\documentclass[twocolumn,english,aps,manuscript,prl,reprint,longbibliography,superscriptaddress,footinbib]{revtex4-1}
\usepackage{graphicx}
\usepackage{epstopdf}

\usepackage[T1]{fontenc}
\usepackage[latin9]{inputenc}
\usepackage{amsmath}
\usepackage{amssymb}
\usepackage[english]{babel}
\usepackage{natbib}
\usepackage{ae}

\usepackage{units}

\makeatletter

\usepackage[usenames,dvipsnames]{color}
 \@ifundefined{textcolor}{}
 {%
   \definecolor{BLACK}{gray}{0}
   \definecolor{WHITE}{gray}{1}
   \definecolor{RED}{rgb}{1,0,0}
   \definecolor{GREEN}{rgb}{0,1,0}
   \definecolor{BLUE}{rgb}{0,0,0.75}
   \definecolor{CYAN}{cmyk}{1,0,0,0}
   \definecolor{MAGENTA}{cmyk}{0,1,0,0}
   \definecolor{YELLOW}{cmyk}{0,0,1,0}
 }
\definecolor{light-gray}{gray}{0.55}

\usepackage{babel}

\usepackage{color}
\usepackage{url}

\usepackage[colorlinks]{hyperref}
\hypersetup{%
        plainpages=true,
        breaklinks=true,
        hypertexnames=false,
        pageanchor=true,
        colorlinks=true,
        linkcolor={blue},
        citecolor={magenta},
        urlcolor={blue},
        anchorcolor={black}
      }
\usepackage[all]{hypcap} 
      
\usepackage{mleftright} 
      
\newcommand{\figref}[1]{\mbox{Fig.~\ref{#1}}}
\newcommand{\tabref}[1]{\mbox{Table~\ref{#1}}}

\renewcommand{\eqref}[1]{\mbox{Eq.~(\ref{#1})}}

\newcommand{\figpanel}[2]{Fig.~\hyperref[#1]{\ref*{#1}(#2)}} 
\newcommand{\figpanels}[3]{Fig.~\hyperref[#1]{\ref*{#1}(#2)-(#3)}} 
\newcommand{\figpanelNoPrefix}[2]{\hyperref[#1]{\ref*{#1}(#2)}} 
\newcommand{\figpanelsNoPrefix}[3]{\hyperref[#1]{\ref*{#1}(#2)-(#3)}} 

\newcommand{\sz}{\sigma_z}

\newcommand{\sm}{\sigma_-}
\renewcommand{\sp}{\sigma_+}
\newcommand{\expec}[1]{\left\langle #1 \right\rangle}
\newcommand{\abs}[1]{\left|#1\right|}
\newcommand{\abssq}[1]{\left| #1 \right|^2}

\DeclareMathAlphabet{\mathpzc}{OT1}{pzc}{m}{it}
\makeatother

\newcommand{\be}{\begin{equation}}
\newcommand{\ee}{\end{equation}}
\newcommand{\bea}{\begin{eqnarray}}
\newcommand{\eea}{\end{eqnarray}}

\AtBeginDocument{%
    \newwrite\bibnotes
    \def\bibnotesext{Notes.bib}
    \immediate\openout\bibnotes=\jobname\bibnotesext
    \immediate\write\bibnotes{@CONTROL{REVTEX41Control}}
    \immediate\write\bibnotes{@CONTROL{%
    apsrev41Control,author="08",editor="1",pages="0",title="0",year="1"}}
     \if@filesw
     \immediate\write\@auxout{\string\citation{apsrev41Control}}%
    \fi
}%

\begin{document}

\title{Deterministic loading and phase shaping of microwaves onto a single artificial atom} 

\author{W.-J.~Lin}
\thanks{These authors contributed equally}
\affiliation{Department of Physics, National Tsing Hua University, Hsinchu 30013, Taiwan}

\author{Y.~Lu}
\email{yongl@chalmers.se}
\thanks{These authors contributed equally}
\affiliation{Department of Microtechnology and Nanoscience (MC2), Chalmers University of Technology, SE-412 96 Gothenburg, Sweden}

\author{ P.~Y.~Wen}
\thanks{These authors contributed equally}
\affiliation{Department of Physics, National Tsing Hua University, Hsinchu 30013, Taiwan}

\author{Y.-T.~Cheng}
\affiliation{Department of Physics, National Tsing Hua University, Hsinchu 30013, Taiwan}

\author{C.-P.~Lee}
\affiliation{Department of Physics, National Tsing Hua University, Hsinchu 30013, Taiwan}

\author{K.-T.~Lin}
\affiliation{CQSE, Department of Physics, National Taiwan University, Taipei 10617, Taiwan}

\author{K.-H.~Chiang}
\affiliation{Department of Physics, National Central University, Jhongli, 32001, Taiwan}

\author{M.~C.~Hsieh}
\affiliation{Department of Physics, National Tsing Hua University, Hsinchu 30013, Taiwan}

\author{J.~C.~Chen}
\affiliation{Department of Physics, National Tsing Hua University, Hsinchu 30013, Taiwan}
\affiliation{Center for Quantum Technology, National Tsing Hua University, Hsinchu 30013, Taiwan}

\author{C.-S.~Chuu}
\affiliation{Department of Physics, National Tsing Hua University, Hsinchu 30013, Taiwan}
\affiliation{Center for Quantum Technology, National Tsing Hua University, Hsinchu 30013, Taiwan}

\author{F.~Nori}
\affiliation{Theoretical Quantum Physics Laboratory, RIKEN Cluster for Pioneering Research, Wako-shi, Saitama 351-0198, Japan}
\affiliation{Physics Department, The University of Michigan, Ann Arbor, Michigan 48109-1040, USA}

\author{A.~F.~Kockum}
\affiliation{Department of Microtechnology and Nanoscience (MC2), Chalmers University of Technology, SE-412 96 Gothenburg, Sweden}

\author{G.-D.~Lin}
\affiliation{CQSE, Department of Physics, National Taiwan University, Taipei 10617, Taiwan}

\author{P.~Delsing}
\affiliation{Department of Microtechnology and Nanoscience (MC2), Chalmers University of Technology, SE-412 96 Gothenburg, Sweden}

\author{I.-C.~Hoi}
\email{ichoi@phys.nthu.edu.tw}
\affiliation{Department of Physics, National Tsing Hua University, Hsinchu 30013, Taiwan}
\affiliation{Center for Quantum Technology, National Tsing Hua University, Hsinchu 30013, Taiwan}

\maketitle


\textbf{Loading quantum information deterministically onto a quantum node is an important step towards a quantum network. Here, we demonstrate that coherent-state microwave photons, with an optimal temporal waveform, can be efficiently loaded onto a single superconducting artificial atom in a semi-infinite one-dimensional (1D) transmission-line waveguide. Using a weak coherent state (average photon number $N \ll 1$) with an exponentially rising waveform, whose time constant matches the decoherence time of the artificial atom, we demonstrate a loading efficiency of above \unit[94]{\%} from 1D semi-free space to the artificial atom. We also show that Fock-state microwave photons can be deterministically loaded with an efficiency of \unit[98.5]{\%}. We further manipulate the phase of the coherent state exciting the atom, enabling coherent control of the loading process. Our results open up promising applications in realizing quantum networks based on waveguide quantum electrodynamics (QED).}


Quantum networks~\cite{Kimble2008}, consisting of quantum nodes and quantum channels, is a topic of intense research, spurred by the vision of a global quantum internet~\cite{Wehner2018}. Quantum nodes can process quantum information, whereas quantum channels can transmit quantum information. The connectivity and scalability of quantum networks strongly depend on the ability to deterministically load quantum information from photons in quantum channels (e.g., free space) onto quantum nodes (e.g., qubits). This loading requires strong interaction between the qubit and the photons, but this is very hard to achieve in 3D free space due to a spatial mode mismatch~\cite{Tey2008}. Attempts have been made using atomic ensembles~\cite{Zhang2012} to enhance the atom-field interaction. However, the loading efficiency only reached \unit[20]{\%}.

Strong interaction between a single artificial atom (a superconducting qubit) and propagating microwave photons has been achieved in a 1D open transmission line~\cite{Astafiev2010, Hoi2011, Gu2017}. This has enabled many important quantum-optical experiments in 1D waveguide QED in superconducting circuits in the past decade~\cite{Gu2017, Roy2017, Astafiev2010, Hoi2011, Hoi2012, VanLoo2013, Hoi2013a, Hoi2015, Forn-Diaz2017, Wen2018, Wen2019, Mirhosseini2019, Wen2020, Kannan2020, Vadiraj2020}. Temporal dynamics has been studied for both a single artificial atom in such a system~\cite{Abdumalikov2011} and for a single real atom in free space~\cite{Leong2016, Aljunid2013, Wang2011}. Impressive progress has been achieved in using a cavity for loading~\cite{Liu2014} (catching) in the optical (microwave~\cite{Wenner2014}) regime, and for quantum state transfer~\cite{Kurpiers2018}. However, deterministic loading of propagating photons onto a single atom in real time, which would be an important component in a quantum network, has not yet been achieved.

\begin{figure}
\includegraphics[width=\linewidth]{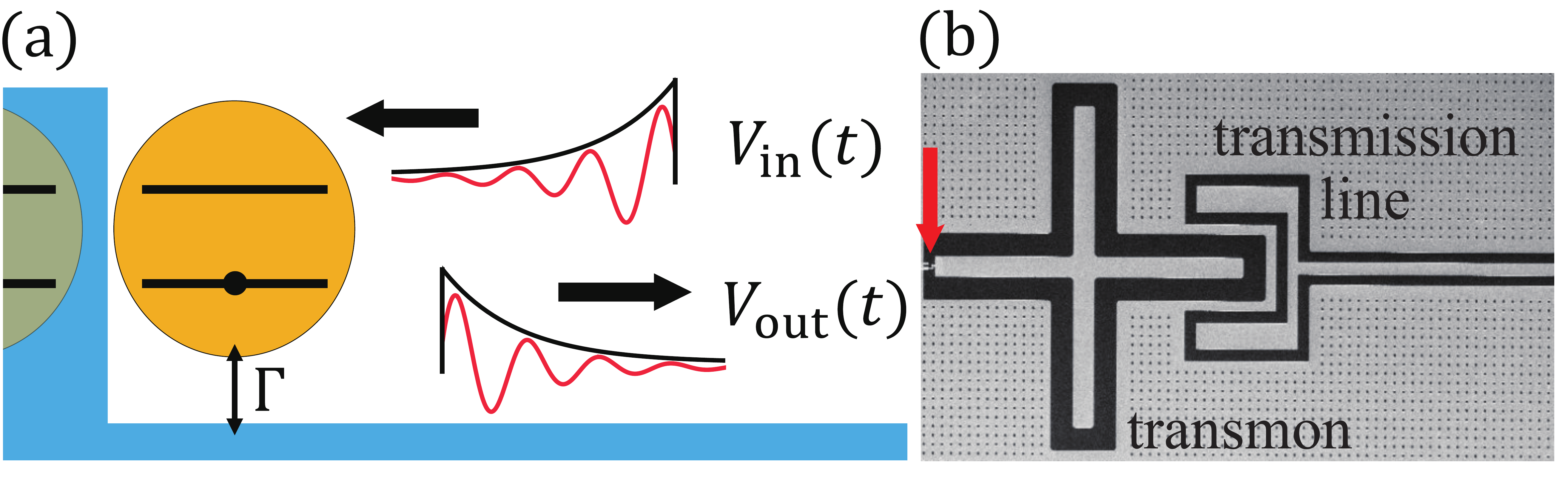}
\caption{(a) Setup sketch. A superconducting artificial atom (yellow) in semi-infinite 1D space, terminated by a mirror. A resonant coherent drive with voltage $V_{\rm in}(t)$ and exponentially rising waveform is sent towards the atom. After interaction with strength $\Gamma$ between the atom and the input field, the atom emits an exponentially decaying output field $V_{\rm out}(t)$. (b) Photo of Sample 2 showing a transmon qubit located at the end of a transmission line, terminated by an open-end capacitor. The transmon contains a SQUID (Superconducting QUantum Interference Device, indicated by the red arrow) loop. Therefore, the atomic resonance frequency is tunable by an external magnetic flux.
\label{fig:Device}}
\end{figure}

In this Letter, we demonstrate that photons in a weak coherent state can be efficiently loaded onto a single artificial atom in semi-infinite 1D free space. Our sample, depicted in \figref{fig:Device}, is a superconducting circuit with a transmon qubit~\cite{Koch2007} coupled to a 1D semi-infinite waveguide, terminated by a mirror. We perform experiments using a weak coherent state with different pulse shapes, including exponentially rising (the time-reversed shape of a photon emitted by decay), exponentially decaying, square, and Gaussian waveforms. When an incident exponentially rising coherent state interacts with the qubit, destructive interference between the atomic emission and the incident field reflecting from the mirror leads to extinction of the output field. This perfect destructive interference occurs when mode matching is achieved. After the pulse is turned off, the atom emits an exponentially decaying field. The loading efficiency is given by the ratio of the coherent output energy and the coherent input energy.

We measure the loading efficiencies for exponentially rising and square pulses. We show that an exponentially rising pulse, whose time constant matches the decoherence time of the artificial atom, gives \unit[12]{\%} higher loading efficiency than the square pulse. In particular, we demonstrate a loading efficiency of \unit[94]{\%} for the exponentially rising pulse with an average photon number $N \ll 1$. Since the two-level atom is highly non-linear, driving it with a high-power coherent state will saturate it, leading to a low loading efficiency. Moreover, we show that a Fock-state photon with the same pulse shape can be deterministically loaded with an efficiency of \unit[98.5]{\%}.

We further manipulate the phase of the coherent input state, enabling coherent control of the loading process~\cite{Specht2009}. By interleaving segments with phases 0 and $\theta$ in the exponentially rising pulse, the rotation axis of the qubit state changes during the excitation process. For $\theta = \pi$, almost no loading will occur. By varying $\theta$, we are thus able to tune how much energy is loaded onto the qubit.


\paragraph*{Frequency-domain characterization.}

We have measured two samples in this work. We first characterize the transmon qubit using single- and two-tone spectroscopy. In single-tone spectroscopy, we measure the reflection coefficient $r$ (magnitude and phase response) of a weak coherent probe tone ($\Omega_{\rm p} \ll \gamma$, where $\Omega_{\rm p}$ is the probe Rabi frequency and $\gamma$ is the decoherence rate of the qubit) versus the probe frequency $\omega_{\rm p}$. Following Ref.~\cite{Probst2015, Lu2019}, we extract, from fitting the data, the qubit transition frequency $\omega_{10}$, the relaxation rate $\Gamma = 1/T_1$, where $T_1$ is the relaxation time, and $\gamma = 1 / T_2$, where $T_2$ is the decoherence time. We also measure the amplitude of the reflection coefficient $\abs{r}$ versus the resonant incident power $P_{\rm in}$, which lets us calibrate the relation between $P_{\rm in}$ and $\Omega_{\rm p}$. The calibration is done by finding the critical power $P_{\rm in} = \hbar \omega_{10} \Gamma / 8$, assuming negligible pure dephasing, where coherent emission is perfectly suppressed, i.e., $\abs{r} = 0$~\cite{Lu2019}. From the two-tone spectroscopy, we extract the transmon charging energy $E_C$. All extracted parameters are summarized in Table~\ref{tab:Parameters}. Note that the rather weak coupling ($\gamma/ 2 \pi\approx \unit[1]{MHz}$) is intentionally chosen to enable resolving the qubit dynamics in the time domain with a nano-second digitizer. Although this coupling is weak compared to many other experiments in superconducting waveguide QED, the qubit-field coupling is still in the strong coupling regime, where the relaxation rate $\Gamma$ is much greater than the pure dephasing rate $\Gamma_\phi$ and non-radiative relaxation. We use all these extracted parameters to simulate the qubit response in the time-domain measurements in the rest of the paper. Further details on the experimental setup and characterization of the two samples are given in~\cite{SupplementaryInformation}.

\begin{table*}
\centering
\begin{tabular}{| c || c | c | c | c | c | c | c | c | c | c |}
\hline
 Sample & $E_C / h$ [MHz] & $E_J / h$ [GHz] & $E_J / E_C$ & $\omega_{10} / 2 \pi$ [GHz] & $\Gamma / 2 \pi$ [MHz] & $\Gamma_\phi / 2 \pi$ [MHz] & $\gamma / 2 \pi$ [MHz] & $T_1 [ns]$ & $T_2 [ns]$  & $\eta$ \\
\hline
 1 & 385 & 8.9 & 23 & 4.8514 & $1.686 \pm 0.007$ & $0.113 \pm 0.009$ & $0.956 \pm 0.005$ & $94.4 \pm 0.4$ & $166 \pm 1$ & $\unit[77.7 \pm 1.6]{\%}$ \\
\hline
 2 & 200 & 15.7 & 78 & 4.8187 & $2.046 \pm 0.003$ & $0.031 \pm 0.004$ & $1.054 \pm 0.003$ & $77.8 \pm 0.1$ & $151 \pm 0.4$ & $\unit[94.2 \pm 0.7]{\%}$ \\
\hline
\end{tabular}
\caption{Extracted and derived qubit parameters for Samples 1 and 2. We extract $\omega_{10}$, $\Gamma$, and $\gamma$ by fitting the magnitude and phase data from single-tone spectroscopy~\cite{Probst2015, Lu2019}. From the two-tone spectroscopy, we extract the anharmonicity, which approximately equals the charging energy $E_C$ of the transmon qubits. We calculate $\Gamma_\phi$ from $\Gamma$ and $\gamma$, using $\gamma = \Gamma / 2 + \Gamma_\phi$. We calculate $E_J$ and $E_J / E_C$ from $\omega_{10}$ and $E_C$, where $\omega_{10} \simeq \sqrt{8 E_J E_C} - E_C$. We calculate $T_1$ and $T_2$ from $\Gamma$ and $\gamma$, respectively. We use a travelling-wave parametric amplifier (TWPA~\cite{Macklin2015}) in Sample 2, which reduces uncertainties compared to Sample 1.
\label{tab:Parameters}}
\end{table*} 


\paragraph*{Loading a coherent state with exponentially rising waveforms onto a qubit.}

We now study the time dynamics of the qubit response to a short pulse. We input an exponentially rising pulse with voltage amplitude
\be
V_{\rm in}(t) = V\Theta \mleft( t_0 - t \mright)e^{\mleft( t - t_0 \mright) / \tau},
\label{EqV}
\ee
where $\Theta$ is the Heaviside step function, $t_0$ is the time when the pulse is turned off, and $\tau$ is the characteristic time of the exponentially rising waveform. Given $V$ and $\tau$, the average photon number $N = \int_0^{t_0} P_{\rm in}(t) dt / (\hbar \omega_{10})$ is fixed; $P_{\rm in}(t) = V_{in}(t)^2 / (2 Z_0)$ is the input power and $Z_0$ is the $\unit[50]{\Omega}$ impedance of the transmission line. For example, $N = 0.09$ for an input power of $\unit[-144]{dBm}$ (the critical power in the single-tone spectroscopy) and $\tau = \unit[145]{ns}$ (close to $T_2$).

\begin{figure*}[ht!]
\includegraphics[width=\linewidth]{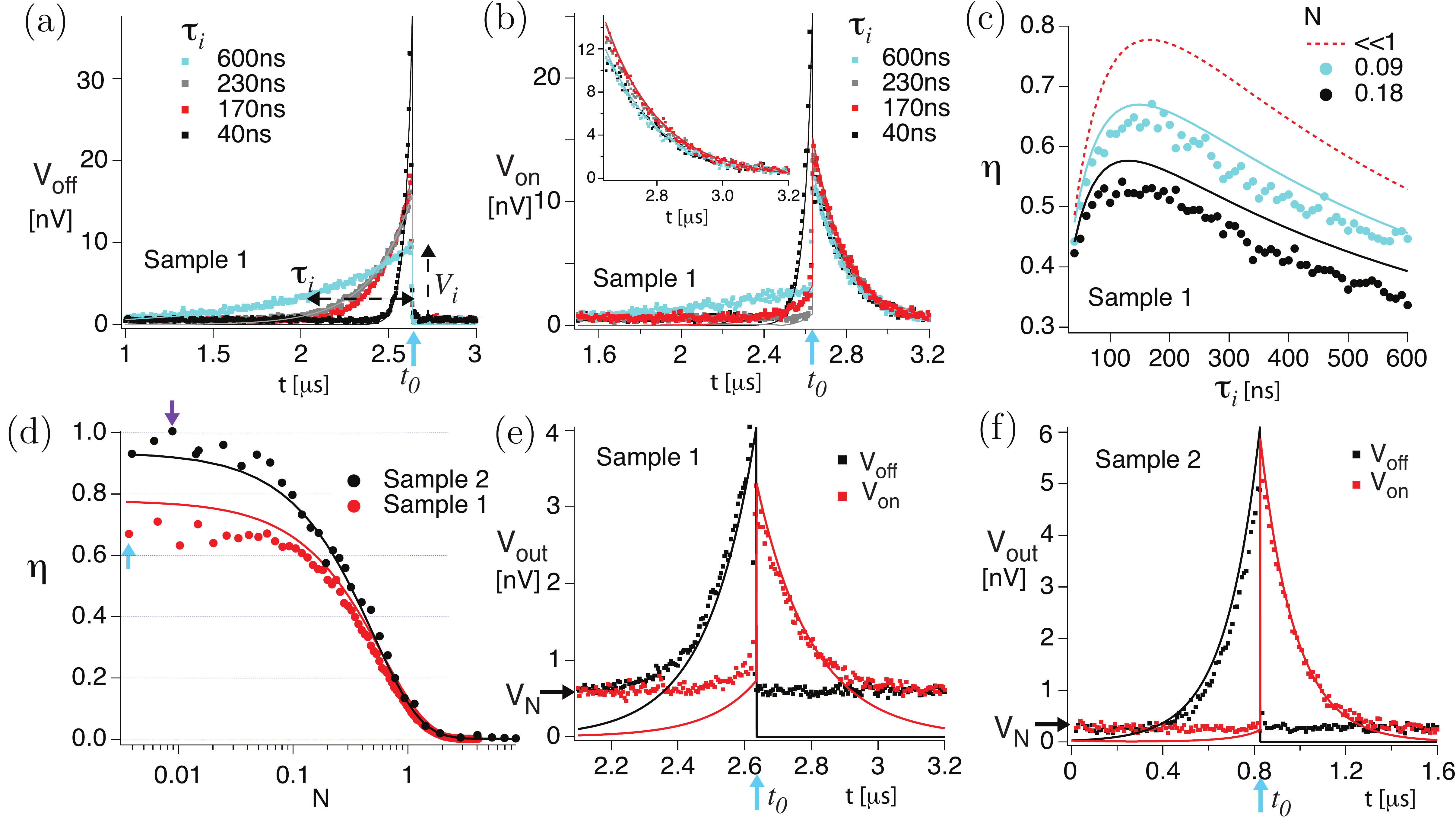}
\caption{Loading a coherent state with exponentially rising waveforms onto a qubit. 
Experimental data are shown as square or round markers. Theoretical calculations, based on parameters in Table~\ref{tab:Parameters} and the equations in the main text, are shown as curves.
(a) (Sample 1) Output magnitude for off-resonant input pulses with four different rise times ($40$, $170$, $230$, and $\unit[600]{ns}$) with constant $N = 0.09$.
(b) (Sample 1) Output magnitude for resonant input, where the qubit first absorbs the input field and then emits a field when the pulse stops at $t_0 = \unit[2.635]{\mu s}$. When the pulse is on, destructive interference between the qubit emission and the reflected input field suppresses the output. Nearly perfect interference between input and qubit emission is observed for $\tau = \unit[230]{ns}$. After the pulse is turned off, the atom emits a coherent field. The amount of emitted field depends on $\tau$. The inset is a zoom-in on the emission. The curves for $\tau_i  = \{ \unit[170]{ns}$, $\unit[230]{ns} \}$ almost overlap, but still differ by \unit[2.4]{\%} loading efficiency.
(c) (Sample 1) Loading efficiency $\eta$ versus $\tau_i$ for different input photon numbers $N = \{ 0.09, 0.18 \}$. The maximum loading efficiency occurs around $\tau_i = T_2$, consistent with the input pulse being the time-reversed version of the output. For higher input power, power broadening of the qubit linewidth causes the maximum loading efficiency to occur at a shorter time. The red dashed curve shows the analytical result from \eqref{EqEta} for $\eta$ versus $\tau$ assuming a weak drive $\Omega \ll \gamma$, i.e., $N \ll 1$.
(d) Loading efficiency versus $N$ for Samples 1 and 2 at fixed $\tau \simeq T_2$. As expected, at $N > 1$, incoherent emission becomes dominant~\cite{Hoi2012}, leading to $\eta \rightarrow 0$. The blue and purple arrows correspond to panels (e) and (f), respectively.
(e, f) Input (black) and emitted (red) voltage at low $N$ for $\tau \simeq T_2$ for Samples 1 and 2, showing the time-reversal symmetry between the input and output fields. At these points $\eta$ is measured to be \unit[67]{\%} and \unit[100.4]{\%} for Samples 1 and 2, respectively. The time resolution for measuring Samples 1 and 2 is $\unit[5]{ns}$ and $\unit[10]{ns}$, respectively. The error in measurement of $\eta$ is mainly from $V_N$ and digitizer resolution.  
\label{fig:Optimal}}
\end{figure*}

We fix $N = 0.09$, and vary the characteristic time $\tau_i$ from $\unit[20]{ns}$ to $\unit[600]{ns}$ (see~\cite{SupplementaryInformation}); the corresponding voltage is
\be
\label{EqVi}
V_i = V \sqrt{\tau / \tau_i} \sqrt{\mleft( 1 - e^{- 2 t_0 / \tau} \mright) / \mleft( 1 - e^{- 2 t_0 / \tau_i} \mright)},
\ee
as shown in \figpanel{fig:Optimal}{a}. Note that the input field is fully reflected by the mirror, such that $V_{\rm off} = V_{\rm in}$, where $V_{\rm off}$ is the average output voltage when the qubit is far detuned from the drive frequency. For resonant excitation, input-output theory gives~\cite{Lalumiere2013}
\be
\label{EqReflection}
\alpha_{\rm out}(t) = \alpha_{\rm in}(t) - i \sqrt{\Gamma} \expec{\sm (t)},
\ee
where $\alpha_{\rm out}$ ($\alpha_{\rm in}$) is the amplitude of the output (input) coherent field in units of $\sqrt{\text{photons/s}}$ and $\sigma_\pm$ is the atomic raising/lowering operator. This gives the Rabi frequency $\Omega(t) = 2 \sqrt{\Gamma} \alpha_{\rm in}(t) = k \sqrt{P_{\rm in}(t)}$. The atom-field coupling constant $k$ is calibrated by frequency-domain characterization (see~\cite{SupplementaryInformation}), which lets us calculate $V_{\rm out}$ at the sample.


The dynamics of the output field is governed by $\expec{\sm (t)}$, which is given by the Bloch equations
\bea
\partial_t \expec{\sigma_\pm} &=& - \gamma \expec{\sigma_\pm} + \Omega(t) \expec{\sz} / 2,  
\label{EqBlochPM}
\\
\partial_t \expec{\sigma_z} &=& - \Gamma \mleft( 1 + \expec{\sz} \mright) - \Omega(t) \mleft( \expec{\sp} + \expec{\sm} \mright),
\label{EqBlochZ}
\eea
where $\sz$ is the third Pauli spin operator.
We numerically solve Eqs.~(\ref{EqBlochPM})--(\ref{EqBlochZ}) with a known arbitrary input waveform $\Omega(t)$ and the parameters in \tabref{tab:Parameters}. All theory curves shown in the whole paper have no free fitting parameters. Since the qubit is initially in the ground state, $\expec{\sigma_\pm(0)} = 0$ and $\expec{\sz(0)} = - 1$. After the pulse stops at $t_0$, the emission decays on the $T_2$ timescale from $\expec{\sm (t_0)}$, since $\alpha_{\rm out} (t) = - i \sqrt{\Gamma} \expec{\sm (t)}$. To obtain the output, we further use the calibrated gain of the amplifiers~\cite{Footnote}, which is about $\unit[68]{dB}$ for Sample 1 and $\unit[99]{dB}$ for Sample 2. Note that in Sample 2, we use a travelling-wave parametric amplifier (TWPA~\cite{Macklin2015}, see~\cite{SupplementaryInformation}). Therefore, the noise level is lower for Sample 2 than for Sample 1.

Figure~\ref{fig:Optimal}(a,b) show the qubit response to off-resonant ($V_{\rm off}$) and on-resonant ($V_{\rm on}$) input, respectively. The 2D plots from which the linecuts in Fig.~\ref{fig:Optimal}(a,b) are taken are shown in~\cite{SupplementaryInformation}. In \figpanel{fig:Optimal}{a}, where the qubit frequency is detuned far away through an external magnetic flux, the output field is assumed to be the reflected input field. Figure~\figpanelNoPrefix{fig:Optimal}{b} consists of two regions: pulse on and pulse off. The first region is the absorption regime and the second is the emission regime. Thus, absorption and emission processes are well separated in time. One can also understand this as a storage process: the photon is converted into a qubit state, which is emitted back as a photon at a later time. The absorption process corresponds to interference between the incoming field $\alpha_{\rm in}(t)$ and the field $- i \sqrt{\Gamma} \expec{\sm(t)}$ emitted from the atom. The emission process corresponds to solely atomic output, which is proportional to $\expec{\sm}$ and therefore decays on the timescale $T_2$.

We define the loading efficiency~\cite{Wenner2014, Liu2014} $\eta = E_{\rm on} / E_{\rm off}$ with
\bea
E_{\rm off} &\sim& \int_{t_i}^{t_0} \mleft[ \abssq{V_{\rm off}(t)} - \abssq{V_N} \mright] dt,
\label{EqEnegryoff}
\\
E_{\rm on} &\sim& \int_{t_0}^{t_f} \mleft[ \abssq{V_{\rm on}(t)} - \abssq{V_N} \mright] dt,
\label{EqEnegryon}
\eea
where $V_N$ is the system voltage noise, $V_{\rm off}$ ($V_{\rm on}$) is the average output voltage when the qubit is far detuned (on resonance), $E_{\rm on}$ ($E_{\rm off}$) is the energy of the emitted (input) coherent state after (before) $t_0$, $t_i$ is the time when the input field is turned on, and $t_f$ is the time when we stop collecting the emitted field. The times $t_i$ and $t_f$ are chosen to be when the signal is equal to the noise level, and $t_0 = \{ \unit[2.635]{\mu s}, \unit[0.825]{\mu s} \}$ for Samples 1 and 2, respectively. Figure~\figpanelNoPrefix{fig:Optimal}{c} shows $\eta$ versus $\tau_i$ for two different photon numbers $N$ for Sample 1. The maximum loading efficiency occurs around $\tau_i = T_2$, consistent with time-reversal symmetry. 

Assuming a weak drive $\Omega \ll \gamma$, i.e, $N \ll 1$, the loading efficiency can be calculated analytically:
\be
\eta \simeq \frac{\Gamma^2}{\gamma \tau \mleft( \gamma + 1 / \tau \mright)^2}.
\label{EqEta}
\ee
This result is shown in the red dashed curve in \figpanel{fig:Optimal}{c}. In \figpanel{fig:Optimal}{d}, we show, for a constant $\tau \simeq T_2$, $\eta$ versus $N$ for Samples 1 and 2. As expected, at large $N \gtrsim 1$, $\eta$ approaches zero. The $\eta$ reaches its maximum at small $N$. This is consistent with a measurement of the reflection amplitude versus resonant power in~\cite{SupplementaryInformation}. Sample 1 deviates a bit from the theory curve when $N \ll 1$, due to a relatively large noise $V_N$. By using the TWPA in Sample 2, that noise is decreased. However, the time resolution of the digitizer used to measure Sample 2 is only $\unit[10]{ns}$, which limits the measurement accuracy, and therefore limits accuracy of the loading efficiency. The details for obtaining \figpanel{fig:Optimal}{d} are shown in ~\cite{SupplementaryInformation}. In Fig.~\ref{fig:Optimal}(e,f), we show the input and emitted signals for points with high $\eta$ (weak drive) from Samples 1 and 2, respectively. We observe the time-reversal symmetry between the input and output field. In this set of measurements, the value of $\eta$ for Samples 1 and 2 is \unit[67]{\%} and \unit[100.4]{\%}, respectively. The efficiency slightly exceeding \unit[100]{\%} is due to uncertainties in measurement, as explained in the caption of \figref{fig:Optimal}.

\paragraph*{Loading a Fock-state photon.}
We also calculate, the loading efficiency that can be achieved for a Fock-state photon in our setup, using the parameters for Sample 2 in \tabref{tab:Parameters}. The details are shown in~\cite{SupplementaryInformation}. The result is $\eta = \unit[98.5]{\%}$, i.e., even higher than what we can achieve for a very weak coherent input signal. If there is no dephasing, the loading efficiency approaches unity both for coherent-state and single-photon Fock-state pumping when perfect destructive interference between the input and the scattered field is realized during the loading process. However, the loading of a coherent state is less robust to dephasing than the loading of a single-photon Fock state. This is because a coherent state is a superposition of multiple Fock states with definite phase relations, which are more easily affected by dephasing, leading to a lower loading efficiency.


\paragraph*{Loading a weak coherent state onto a qubit with other waveforms.}

\begin{figure*}[ht!]
\includegraphics[width=\linewidth]{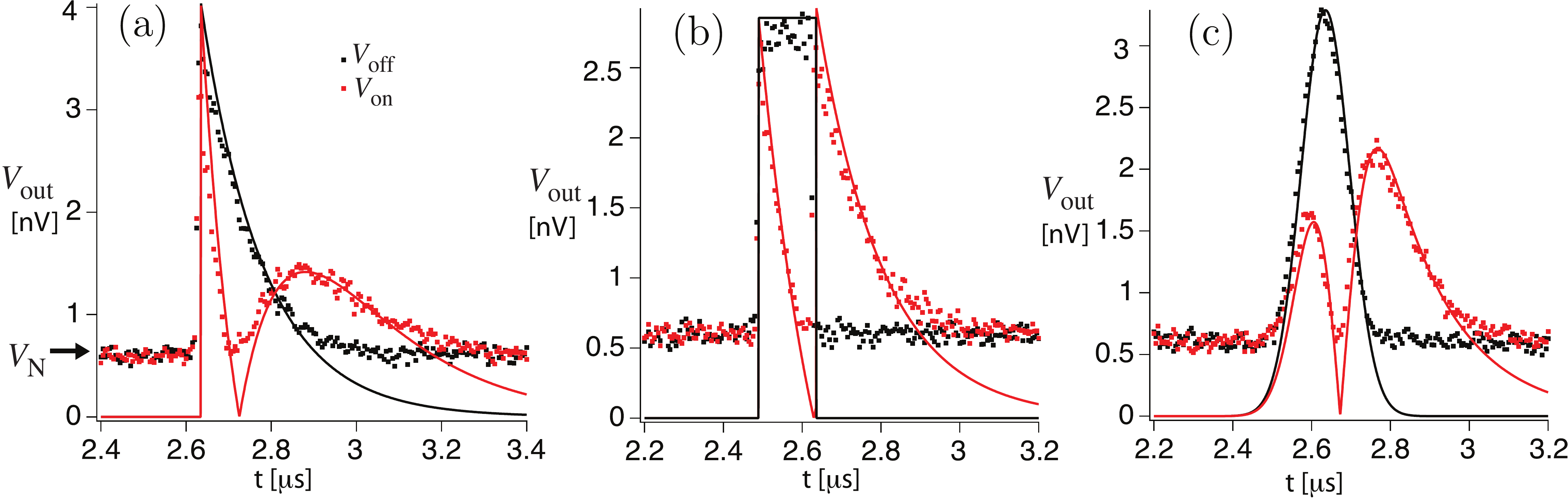}
\caption{Loading a weak coherent state onto the qubit of Sample 1 with other pulse shapes, all with $N = 0.004$. 
In all plots, black and red represents input (off-resonant) and output (resonant), respectively. Experimental data are shown as square markers; each trace is averaged over 450 million runs. Solid curves are theoretical calculations.
(a) Exponentially decaying pulse with characteristic time $\unit[145]{ns}$.
(b) Square pulse with width $\unit[145]{ns}$.
(c) Gaussian pulse with full width at half maximum $\unit[145]{ns}$. 
\label{fig:shape}}
\end{figure*}

For comparison, we study different input pulse shapes with the same $N$ for Sample 1: exponentially decaying, square, and Gaussian. In \figpanel{fig:shape}{a}, we show the exponentially decaying waveform, whose power spectrum is the same as for the exponentially rising pulse. Both absorption and emission occur during this pulse, which leads to waveform distortion. For the square pulse in \figpanel{fig:shape}{b}, the loading efficiency is \unit[55]{\%}, which is \unit[12]{\%} lower than for the exponentially rising pulse. For the Gaussian pulse in \figpanel{fig:shape}{c}, the output splits into two peaks. For exponentially decaying waveform and Gaussian waveform, since there is no clear time separation between absorption and emission, the loading efficiency cannot be well defined.


\paragraph*{Phase shaping of the exponentially rising pulse.}

\begin{figure}[ht!]
\includegraphics[width=\linewidth]{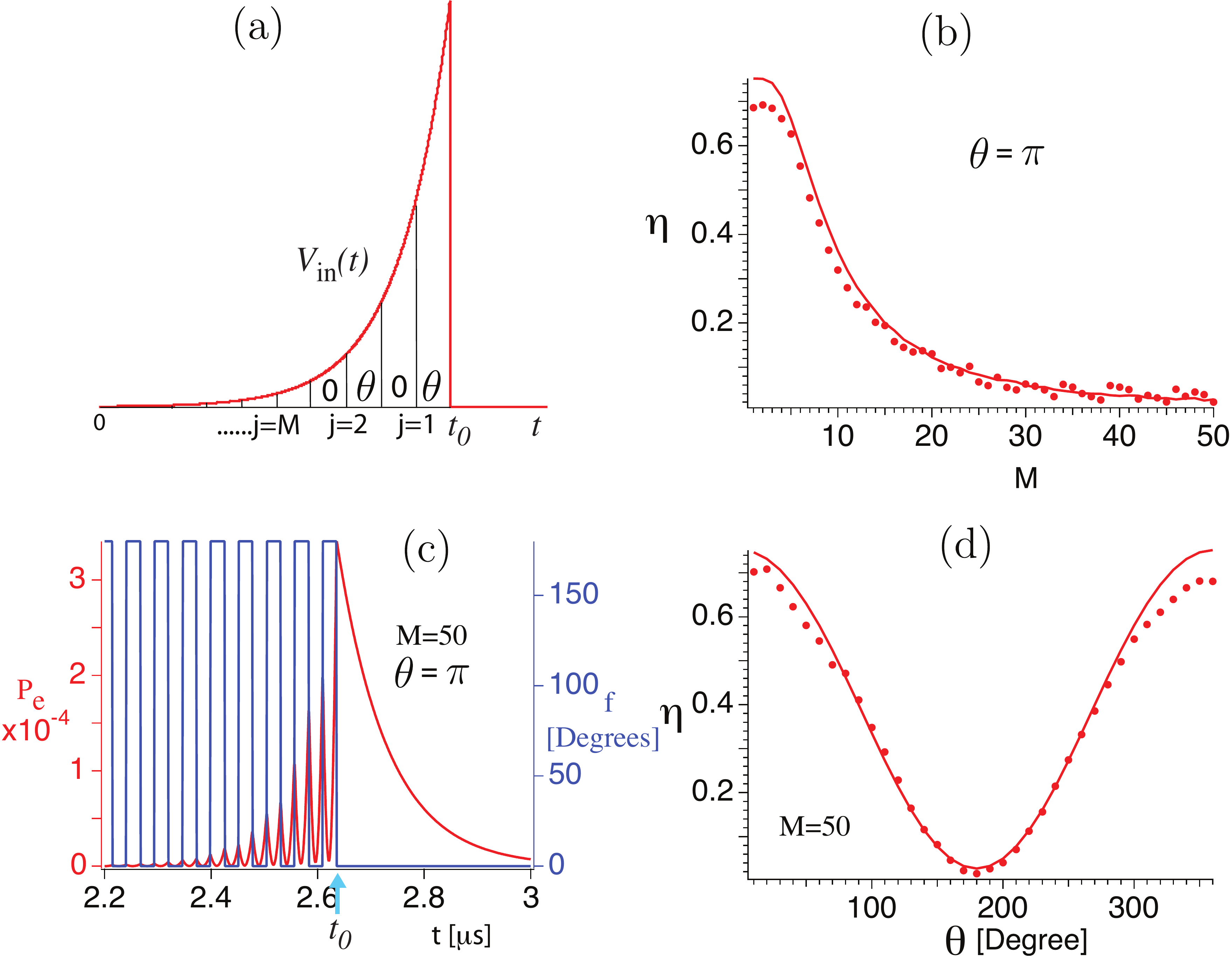}
\caption{Phase shaping of an exponentially rising pulse with $N = 0.015$ for Sample 1.
Experimental data are shown by round markers. Solid curves are theory calculations based on the parameters in \tabref{tab:Parameters} and the expressions in Eqs.~(\ref{EqBlochPM}), (\ref{EqBlochZ}), (\ref{CmplxRabi}), and (\ref{phaseMod}). 
(a) Input exponentially rising pulse with phase shaping alternating between 0 and arbitrary $\theta$ in $M$ intervals between $t = 0$ and $t = t_0$.
(b) Loading efficiency versus $M$ for $\theta = \pi$. (c) Qubit population $P_e$ versus time for $M = 50$, $\theta = \pi$. After a small time segment of excitation, the qubit returns back towards the ground state, leading to low efficiency.
(d) Efficiency with $M = 50$ versus $\theta$. At $\theta = 180$ degrees, the efficiency is close to zero. Maximum efficiency is found for $\theta = \{ 0, 360 \}$ degrees.
\label{fig:interference}}
\end{figure}

All measurements so far were at a constant input phase. Together with waveform control of coherent-state photons, phase shaping not only achieves complete control of single-photon wave packets, but also finds applications in quantum computation~\cite{Knill2001}, where two-photon interference with appropriate phase control is essential. Here, we use phase shaping to control the qubit excitation, and thus the loading efficiency. When we apply phase shaping, the Rabi frequency in Eqs.~(\ref{EqBlochPM},\ref{EqBlochZ}) becomes complex: 
\be
\Omega (\theta, t) = e^{i f(\theta, t)} \Omega(t),
\label{CmplxRabi}
\ee
with
\be
\label{phaseMod}
f (\theta, t) =
\begin{cases}
\theta & j t_M > t \geq (j - 1/2) t_M \\
0 & (j - 1/2) t_M > t \geq (j - 1) t_M,
\end{cases}
\ee
where $j = 1, 2, \ldots, M$, $t_M = t_0 / M$, and $M$ is the number of periods. This is illustrated in \figpanel{fig:interference}{a}.

In \figpanel{fig:interference}{b}, we show $\eta$ versus $M$ for $\theta = \pi$. The loading efficiency decreases with increasing $M$. When $M$ is large, the time segment $t_M$ is small, leading to neighbouring pulse segments having approximately equal area. Since neighbouring pulses have a $\pi$ phase shift, their effects on the qubit will cancel and in the end, the qubit is almost not excited [see \figpanel{fig:interference}{c} for $M = 50$], leading to low emission from the qubit. For details, see~\cite{SupplementaryInformation}. The situation is different when $M$ is small. In that case, the neighbouring pulse segments have different areas, so their effects do not cancel and their net effect is to excite the qubit. This leads to more emission from the qubit and a higher loading efficiency. $M=0$ corresponds to no phase modulation. 

Next, we keep $M = 50$ and vary $\theta$ in \figpanel{fig:interference}{d}. The detailed 2D plots underlying \figpanel{fig:interference}{d} are shown in~\cite{SupplementaryInformation}. We see in \figpanel{fig:interference}{d} that the loading efficiency oscillates versus $\theta$. This is because the phase $\theta$ determines the rotation axis of the qubit state on the Bloch sphere. Depending on the phase, the pulse segments with non-zero $\theta$ will either excite the qubit further or move the qubit back towards the ground state.




\paragraph*{Conclusion and outlook.}

We demonstrate the efficient loading of a weak coherent state onto a qubit in a 1D semi-open waveguide. We obtain loading efficiencies above \unit[94]{\%} using weak exponentially rising coherent input pulses with characteristic time equal to the qubit decoherence time. Furthermore, we calculate that our setup with a qubit in front of a mirror can also be loaded with a Fock-state photon with deterministic efficiency \unit[98.5]{\%} using the parameters measured for Sample 2. The loading efficiency of Sample 2 is better than that of Sample 1 due to a lower pure dephasing rate and a higher relaxation rate. As shown by \eqref{EqEta}, for optimal mode matching $1 / \tau = \gamma$, the loading efficiency $\eta = (1 + 2 \Gamma_{\phi} / \Gamma)^{-2}$, so unit loading efficiency is obtained when $\Gamma \gg \Gamma_\phi$. Sample 2 achieved this by having a high $E_J / E_C$ ratio, which reduced charge noise, and thus the pure dephasing. In conclusion, our results may enable promising applications by realizing deterministic quantum networks based on waveguide quantum electrodynamics. A next step in this direction would be to make the coupling between the qubit and the waveguide tunable to prevent the photon being emitted immediately after it has been absorbed. We note that such a tunable coupling could be achieved in the current setup by placing the qubit close to a node of the field in the waveguide and change the effective distance to the mirror by tuning the qubit frequency~\cite{Hoi2015} or changing the boundary condition~\cite{Forn2017}.


\bibliography{LoadingPhotonRefs}

\begin{thebibliography}{36}
\bibitem [{Foo()}]{Footnote}%
  \BibitemOpen
  \href@noop {} {}\bibinfo {note} {We know the absolute power at the critical
  point at the sample, and we know the power at the critical point at the
  digitizer at room temperature, so we can calibrate the gain.}\BibitemShut
  {Stop}%
\end{thebibliography}

\textbf{Acknowledgments:} I.-C.H.~and J.C.C.~would like to thank I.A. Yu for fruitful discussions. We also acknowledge IARPA and Lincoln Labs for providing the TWPA used in this experiment. This work was financially supported by the Center for Quantum Technology from the Featured Areas Research Center Program within the framework of the Higher Education Sprout Project by the Ministry of Education (MOE) in Taiwan. I.-C.H.~acknowledges financial support from the MOST of Taiwan under project 109-2636-M-007-007. G.-D.L. acknowledges support from MOST, Taiwan, under grant No.~109-2112-M-002-022, and NTU under grant No.~109L892006. A.F.K. and P.D.~acknowledge support from the Knut and Alice Wallenberg Foundation through the Wallenberg Centre for Quantum Technology (WACQT). A.F.K.~also acknowledges support from the Swedish Research Council (grant number 2019-03696). F.N. is supported in part by: NTT Research, Army Research Office (ARO) (Grant No.~W911NF-18-1-0358), Japan Science and Technology Agency (JST) (via the Q-LEAP program and CREST Grant No.~JPMJCR1676), Japan Society for the Promotion of Science (JSPS) (via the KAKENHI Grant No.~JP20H00134 and the JSPS-RFBR Grant No.~JPJSBP120194828), the Asian Office of Aerospace Research and Development (AOARD), and the Foundational Questions Institute Fund (FQXi) via Grant No.~FQXi-IAF19-06.


\textbf{Author Contributions:} I.-C.H.~designed and supervised the experiment. W.-J.L., Y.L., and P.Y.W.~performed the experiments. Y.-T.C.~performed the theory simulations. Y.L.~and C.-P.L.~fabricated the samples. A.F.K., K.-T.L., and G.-D.L. provided theoretical assistance. I.-C.H., W.J.L., and A.F.K. wrote the manuscript with input from all authors.

\end{document}


\title{Supplementary Information for ``Deterministic loading and phase shaping of microwaves onto a single artificial atom''}

\author{W.-J.~Lin}
\thanks{These authors contributed equally}
\affiliation{Department of Physics, National Tsing Hua University, Hsinchu 30013, Taiwan}

\author{Y.~Lu}
\email{yongl@chalmers.se}
\thanks{These authors contributed equally}
\affiliation{Department of Microtechnology and Nanoscience (MC2), Chalmers University of Technology, SE-412 96 Gothenburg, Sweden}

\author{ P.~Y.~Wen}
\thanks{These authors contributed equally}
\affiliation{Department of Physics, National Tsing Hua University, Hsinchu 30013, Taiwan}

\author{Y.-T.~Cheng}
\affiliation{Department of Physics, National Tsing Hua University, Hsinchu 30013, Taiwan}

\author{C.-P.~Lee}
\affiliation{Department of Physics, National Tsing Hua University, Hsinchu 30013, Taiwan}

\author{K.-T.~Lin}
\affiliation{CQSE, Department of Physics, National Taiwan University, Taipei 10617, Taiwan}

\author{K.-H.~Chiang}
\affiliation{Department of Physics, National Central University, Jhongli, 32001, Taiwan}

\author{M.~C.~Hsieh}
\affiliation{Department of Physics, National Tsing Hua University, Hsinchu 30013, Taiwan}

\author{J.~C.~Chen}
\affiliation{Department of Physics, National Tsing Hua University, Hsinchu 30013, Taiwan}
\affiliation{Center for Quantum Technology, National Tsing Hua University, Hsinchu 30013, Taiwan}

\author{C.-S.~Chuu}
\affiliation{Department of Physics, National Tsing Hua University, Hsinchu 30013, Taiwan}
\affiliation{Center for Quantum Technology, National Tsing Hua University, Hsinchu 30013, Taiwan}

\author{F.~Nori}
\affiliation{Theoretical Quantum Physics Laboratory, RIKEN Cluster for Pioneering Research, Wako-shi, Saitama 351-0198, Japan}
\affiliation{Physics Department, The University of Michigan, Ann Arbor, Michigan 48109-1040, USA}

\author{A.~F.~Kockum}
\affiliation{Department of Microtechnology and Nanoscience (MC2), Chalmers University of Technology, SE-412 96 Gothenburg, Sweden}

\author{G.-D.~Lin}
\affiliation{CQSE, Department of Physics, National Taiwan University, Taipei 10617, Taiwan}

\author{P.~Delsing}
\affiliation{Department of Microtechnology and Nanoscience (MC2), Chalmers University of Technology, SE-412 96 Gothenburg, Sweden}

\author{I.-C.~Hoi}
\email{ichoi@phys.nthu.edu.tw}
\affiliation{Department of Physics, National Tsing Hua University, Hsinchu 30013, Taiwan}
\affiliation{Center for Quantum Technology, National Tsing Hua University, Hsinchu 30013, Taiwan}

\date{\today}


\maketitle

\tableofcontents

\renewcommand{\thefigure}{S\arabic{figure}}
\renewcommand{\thesection}{S\arabic{section}}
\renewcommand{\theequation}{S\arabic{equation}}
\renewcommand{\thetable}{S\arabic{table}}
\renewcommand{\bibnumfmt}[1]{[S#1]}
\renewcommand{\citenumfont}[1]{S#1}


\section{Experimental setup and device}

\begin{figure}
\includegraphics[width=\linewidth]{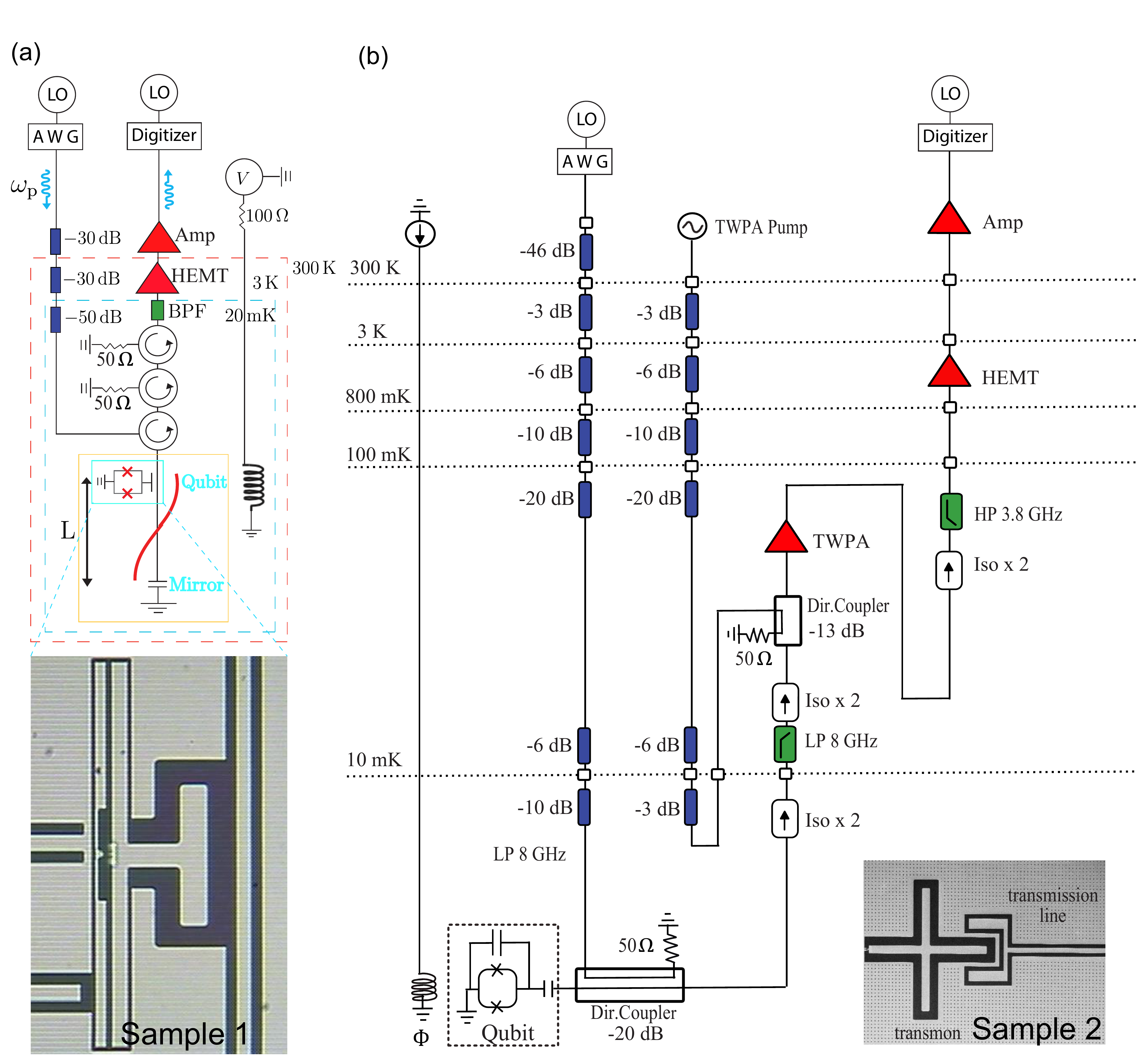}
\caption{Photos of our devices and sketches of the experimental setups.
(a) Measurement setup for Sample 1. In the device image, shown in the zoom-in at the bottom, the two long bright parts form the qubit capacitance, and the extended pad from the one on the right determines the relatively weak capacitive coupling to the transmission line. Two Josephson junctions form a Superconducting QUantum Interference Device (SQUID) loop between the two islands forming the qubit capacitance. In the sketch of the setup, the red curve represents the electromagnetic field distribution along the transmission line. The qubit is placed $L \simeq \unit[12]{mm}$ from the mirror (capacitance to ground at the end of the transmission line). For the qubit frequency $\omega_{10} / 2 \pi = \unit[4.85]{GHz}$, the location corresponds to an antinode of the electromagnetic field. Since the roundtrip time from the qubit to the mirror and back, $2 L / v \sim \unit[0.2]{ns}$, is small compared to the timescale of the atomic evolution, $T_2$, we have neglected the phase term $e^{i \phi}$ acquired by the round trip, which does not affect the dynamics. A superconducting coil controlled by a dc voltage $V$ induces a global magnetic field and enables us to tune the qubit frequency $\omega_{10}$. For measurements, coherent signals at frequency $\omega_p$ are synthesized at room temperature and fed through attenuators (blue rectangles) to the sample, which resides in a cryostat cooled to \unit[20]{mK} to avoid thermal fluctuations affecting the experiment. The reflected signal passes circulators, filters (green rectangle), and amplifiers (red triangles), and is then collected by the digitizer.
(b) Measurement setup of Sample 2. In the device image, shown in the zoom-in in the lower right, the bright cross forms the qubit capacitance. A SQUID loop with two Josephson junctions connects to the left piece of the cross, while the transmission line couples to the qubit through the right piece of the cross. The measurement scheme is similar to (a) except for the inclusion of a travelling-wave parametric amplifier (TWPA), which leads to a noise level $V_N$ about two times smaller than in Sample 1 and smaller uncertainty in extracted qubit parameters.
\label{fig:Fig_Device}}
\end{figure}

Figure~\ref{fig:Fig_Device} shows the experimental setups for Samples 1 and 2. In both samples, transmon qubits are coupled to a semi-infinite one-dimensional (1D) transmission line with characteristic impedance $Z_0 \simeq \unit[50]{\Omega}$. Arbitrary waveform generators (AWG) shape the waveform of input coherent photons, while digitizers capture the reflected signal, enabling us to resolve the time dynamics of the qubit and the photons.    

\section{Steady-state reflection coefficient for extracting parameters of the qubits}

\begin{figure}
\includegraphics[width=\linewidth]{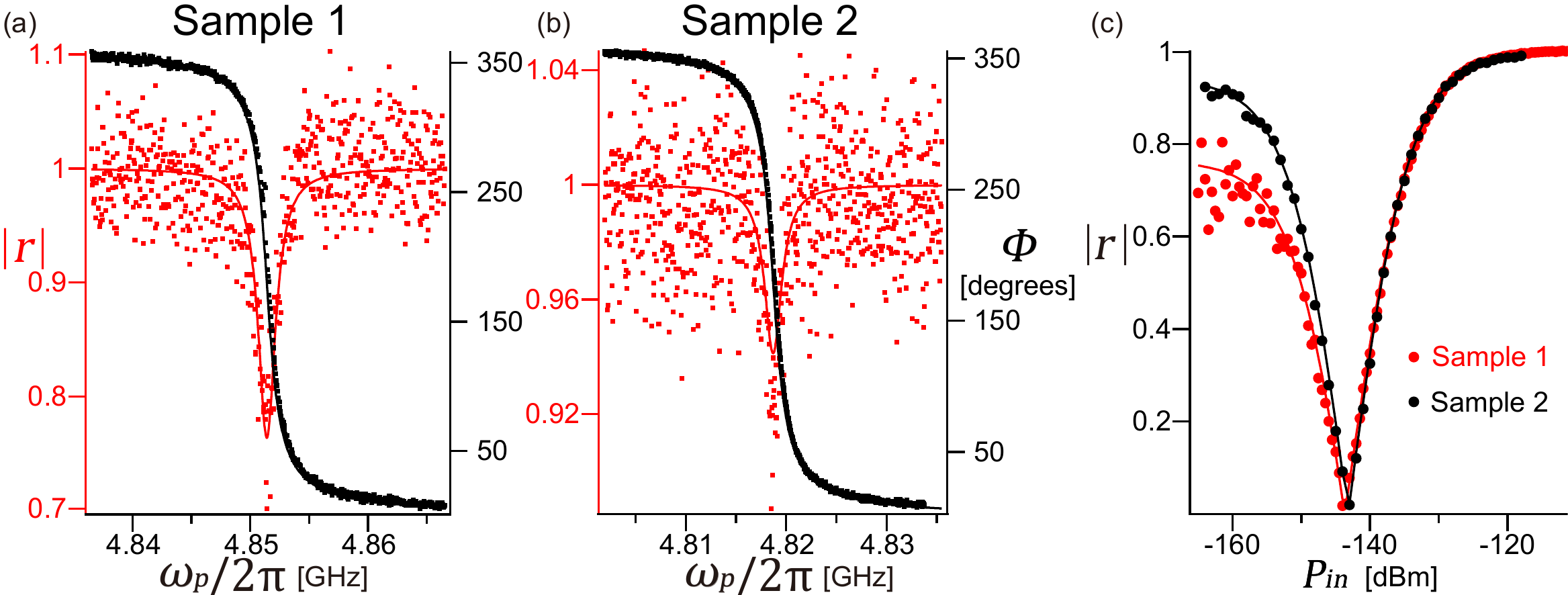}
\caption{Steady-state reflection measurement with a continuous coherent probe.
(a, b) Measured magnitude $\abs{r}$ (red dots) and phase $\Phi$ (black dots) of $r = \abs{r} e^{i \Phi}$ as a function of probe frequency $\omega_p$ with a probe power of $\unit[-166]{dBm}$ for Samples 1 and 2, respectively. Theoretical fits (solid curves) give us the values of $\Gamma$, $\gamma$, and $\omega_{10}$ mentioned in the main text.
(c) Measured magnitude response $\abs{r}$ as a function of the incident resonant power $P_{\rm in}$ for Samples 1 (red dots) and 2 (black dots). The theoretical fits (solid curves) yield the atom-field coupling $k$ mentioned in the main text.
\label{fig:ParametersExtraction}}
\end{figure}

In this section, we describe how we extract the parameters of the qubit from data obtained in the steady-state reflection measurements. The reflection coefficient for a continuous weak probe is given by~\cite{Hoi2015}
\be
r = 1 - \frac{\Gamma}{\gamma + i \delta \omega_p},
\label{eq:weakprobe}
\ee
where $\Gamma$ is the qubit relaxation rate, $\gamma$ is the qubit decoherence rate, and $\delta \omega_p = \omega_{10} - \omega_p$. We apply a weak probe with a power of $\unit[-166]{dBm}$ and measure the reflection coefficient of the two samples. The results are shown in \figpanels{fig:ParametersExtraction}{a}{b}. The solid curves (both magnitude and phase) in \figpanels{fig:ParametersExtraction}{a}{b} are fits to the data using \eqref{eq:weakprobe} and the circle fit technique in~\cite{Lu2019}. This allows us to extract values for $\Gamma$, $\gamma$, and $\omega_{10}$.

For a resonant probe, the reflection coefficient is given by~\cite{Hoi2015}
\be
r = 1 - \frac{\Gamma^2}{\Gamma \gamma + \Omega_p^2},
\label{eq:powerdependent}
\ee
where $\Omega_p$ is the Rabi frequency of the probe. In \figpanel{fig:ParametersExtraction}{c}, we show the measured magnitude of the reflection coefficient as a function of resonant input power $P_{\rm in}$. The nonlinear power dependence allows us to extract $k$ in $\Omega_p = k \sqrt{P_{\rm in}}$ using \eqref{eq:powerdependent}.

\section{Full data for loading coherent photons with exponentially rising waveforms onto a qubit}

\begin{figure}
\includegraphics[width=\linewidth]{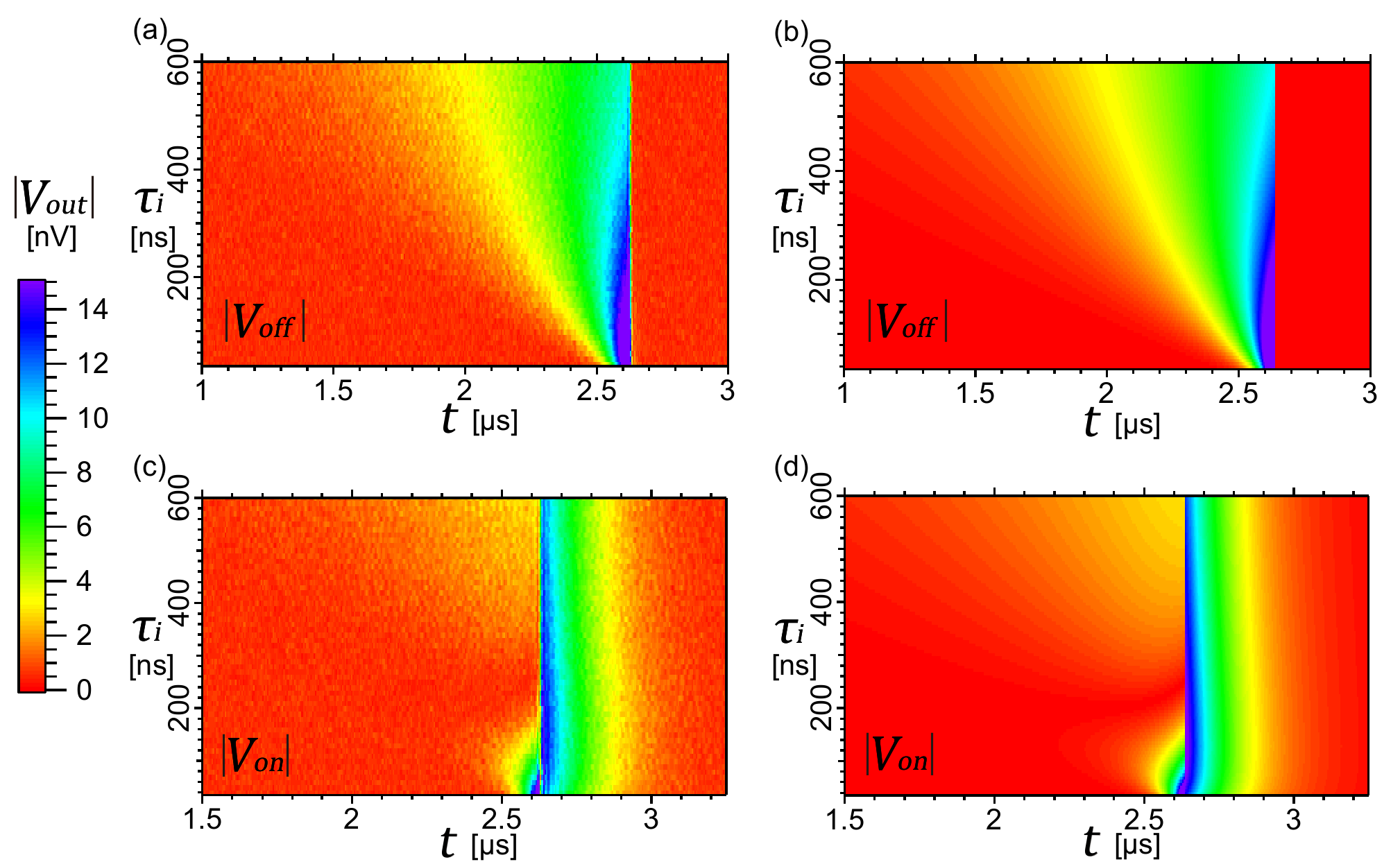}
\caption{Magnitude of the reflected field for Sample 1 when driven with an exponentially rising input coherent pulse with average photon number $N = 0.09$.
(a, c) The experimental data of the reflected field with the qubit far off resonance, $\abs{V_{\rm off}}$, and on resonance, $\abs{V_{\rm on}}$, respectively.
(b, d) Theory simulations of the data in panels (a) and (c), respectively. The simulations are performed using the parameters extracted from \figref{fig:ParametersExtraction} (no free parameters). The agreement between the data and the simulations is excellent. 
\label{fig:FullDataFixedPhoNum}}
\end{figure}

\begin{figure}
\includegraphics[width=\linewidth]{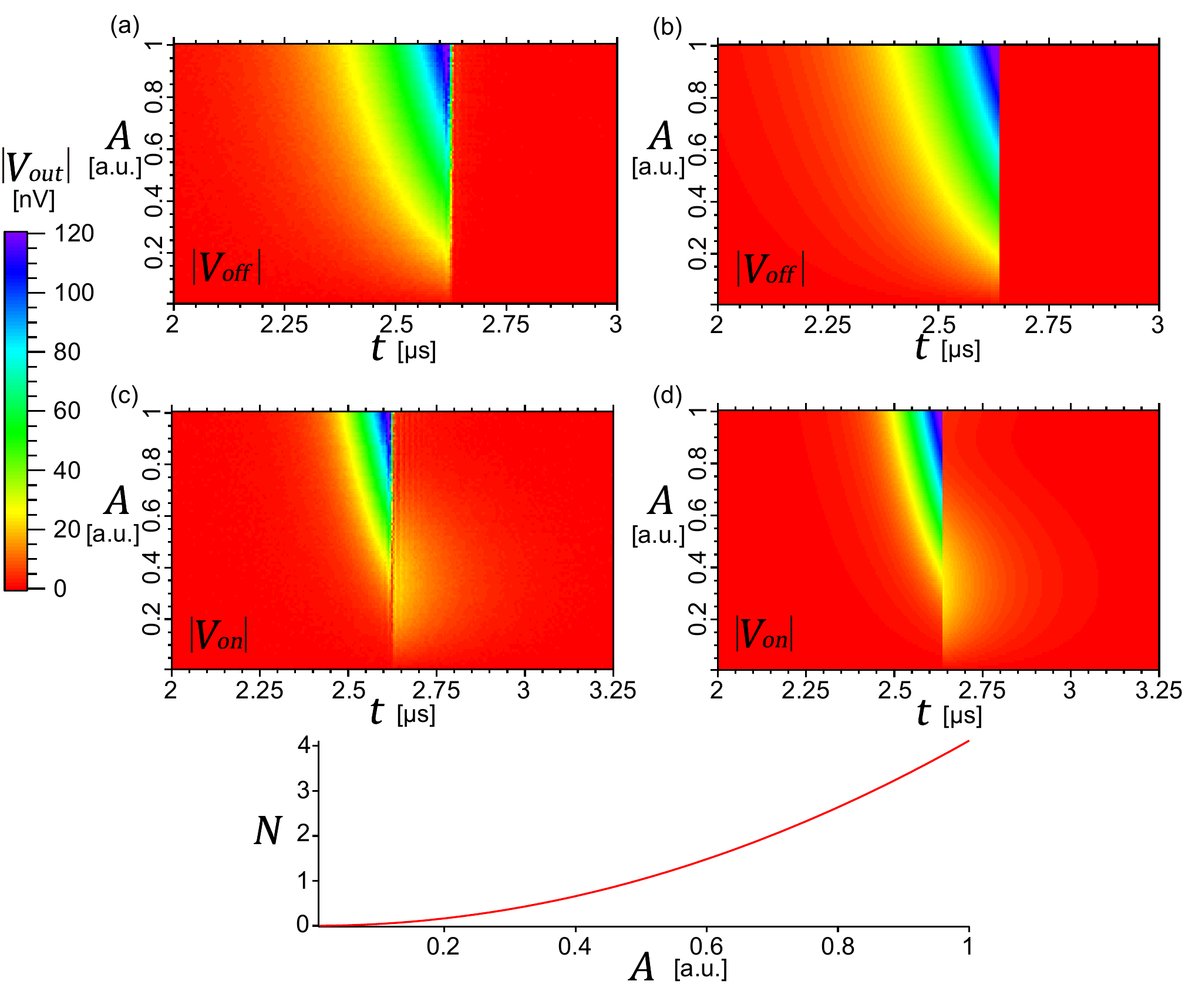}
\caption{Magnitude of the reflected field for Sample 1 when driven with an exponentially rising input coherent pulse with different average photon numbers $N$, but with the same characteristic time $\tau = \unit[145]{ns}$.
(a, c) The experimental data of the reflected field with the qubit far off resonance, $\abs{V_{\rm off}}$, and on resonance, $\abs{V_{\rm on}}$, respectively.
(b, d) Theory simulations of the data in panels (a) and (c), respectively. The simulations are performed using the parameters extracted from \figref{fig:ParametersExtraction} (no free parameters). The agreement between the data and the simulations is excellent.
(e) The relation between average photon number $N$ and amplitude $A$ in panels (a)-(d), where $A$ is given in arbitrary units.
\label{fig:FullDataAmpDependent}}
\end{figure}

In this section, we present the full data of Fig.~2 in the main text. This is the data from which we calculate the loading efficiency $\eta$ in Fig.~2(c)-(d). Figure~\ref{fig:FullDataFixedPhoNum} shows the magnitude of the reflected field as a function of characteristic time $\tau_i$ and time $t$ when the input field has average photon number $N = 0.09$. Figure~\ref{fig:FullDataAmpDependent} shows the magnitude of the reflected field for different average photon numbers $N$ from 0.0004 to 4.12, but with the same characteristic time $\tau$ of \unit[145]{ns}.

\section{Full data of phase-shaping experiments}

\begin{figure}
\includegraphics[width=\linewidth]{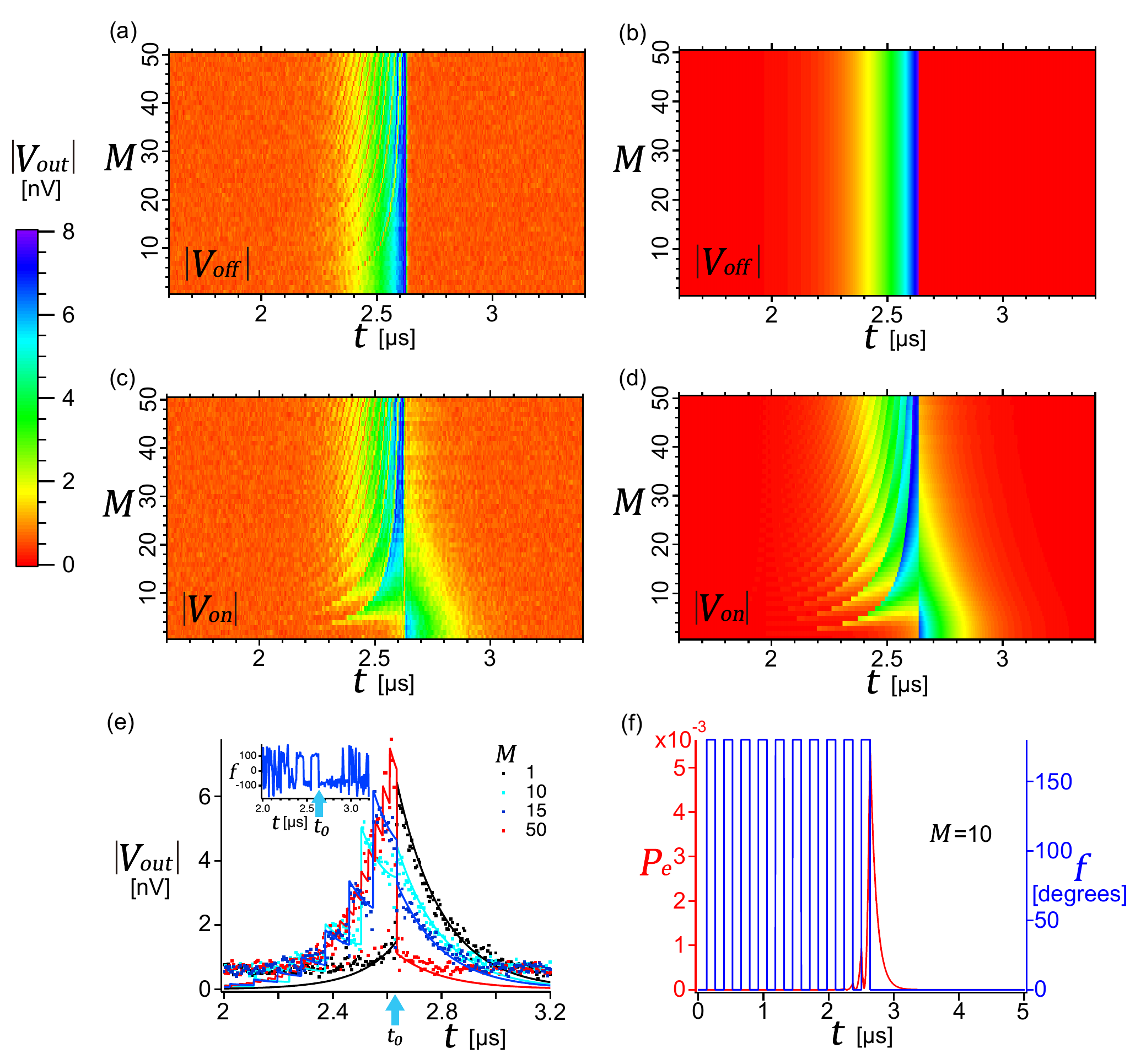}
\caption{Magnitude of the reflected field for Sample 1 with a phase-shaped exponentially rising input coherent pulse with $\theta = \pi$ and varying number of time segments $M$ from 1 to 50.
(a, c) The experimental data of the reflected field with the qubit far off resonance, $\abs{V_{\rm off}}$, and on resonance, $\abs{V_{\rm on}}$, respectively. The time resolution of our digitizer is \unit[5]{ns}, meaning that it records one measurement point by averaging the measured voltage within this short time bin. We observe that there are several measured points with rather low magnitude when phase switching events happen [see also (e)], resulting in the fringes seen here. Within the time bin, the digitizer sometimes measures $+I$, and sometimes $-I$ ($\pi$ phase shift). After averaging, it will be zero, resulting in low magnitude.
(b, d) Theory simulations of the data in panels (a) and (c), respectively. The simulations are performed using the parameters extracted from \figref{fig:ParametersExtraction} (no free parameters). The orange background in (a) and (c) indicates a system noise $V_N = \unit[0.6]{nV}$.
(e) Linecuts of panels (c) and (d) with $M = \{1, 10, 15, 50 \}$.
Inset: the measured phase response for $M = 15$. The time resolution of the digitizer is \unit[5]{ns}, which is too large to resolve the dynamics for $M = 50$ during excitation. The interference occurs between phase-switching events for $t < t_0$, which is consistent with the measured $\pi$ phase shift shown in the inset. When the signal is close to or below the noise floor, the phase is not well defined. We also see phase-shift switching for $t < t_0$, in contrast to constant phase after emission ($t > t_0$). Finally, $M = 1$ gives the highest emission.
(f) Theory plots for the time dynamics of the excited-state population $P_e$ (red curve) and the phase  $f$ of the coherent input (blue curve) with $M = 10$. In the case $M = 50$, mentioned in the Fig.~4(c) of the main text, $P_e$ is smaller than for $M = 10$ here. This indicates that the qubit almost stays completely in its ground state for high $M$, which leads to low emission.
\label{fig:FullDataSlicesNum}}
\end{figure}

\begin{figure}
\includegraphics[width=\linewidth]{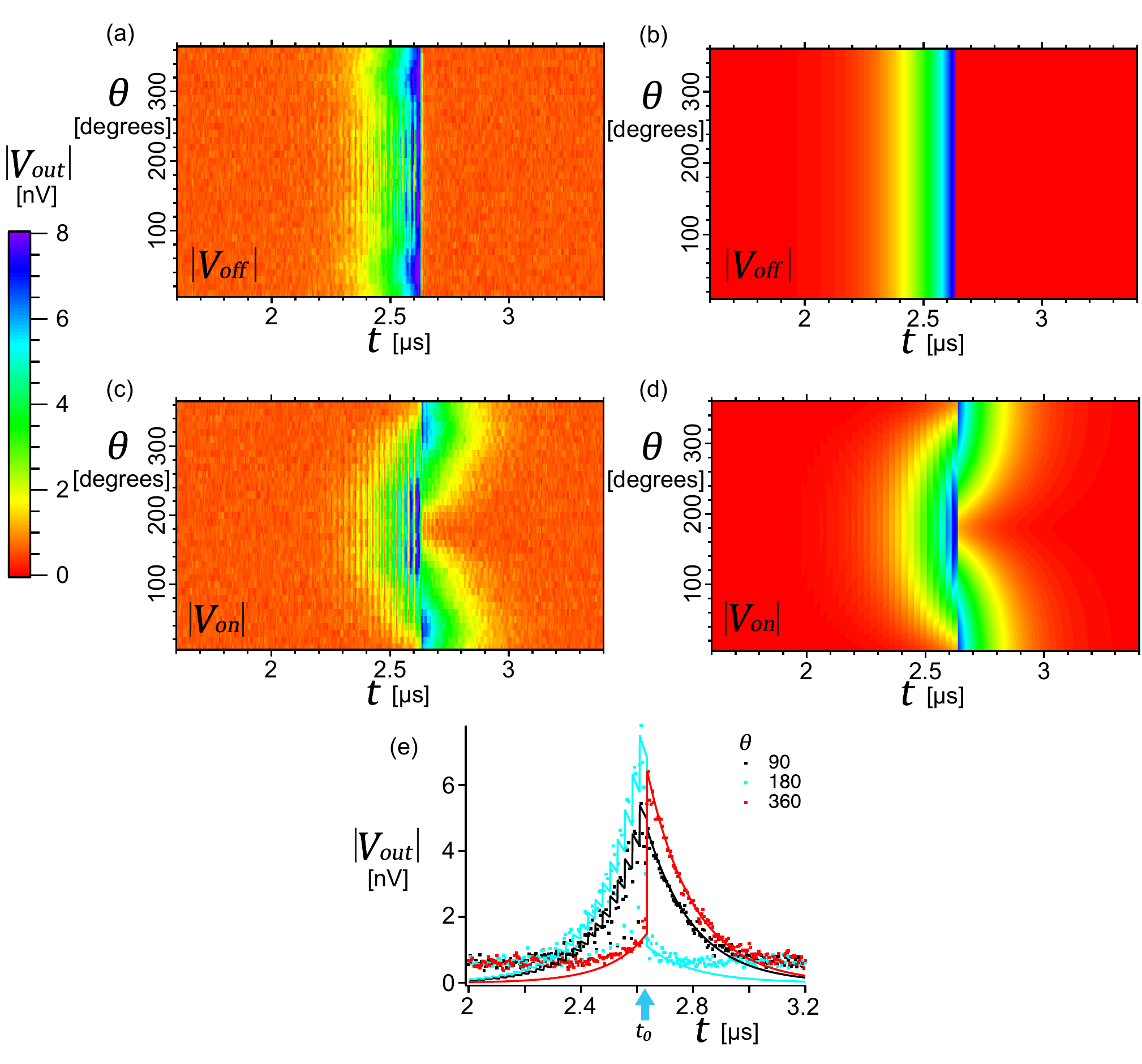}
\caption{Magnitude of the reflected field for Sample 1 with a phase-shaped exponentially rising input coherent pulse with $M = 50$ and  $\theta$ varying between 0 and $2 \pi$.
(a, c) The experimental data of the reflected field with the qubit far off resonance, $\abs{V_{\rm off}}$, and on resonance, $\abs{V_{\rm on}}$, respectively.
(b, d) Theory simulations of the data in panels (a) and (c), respectively. The simulations are performed using the parameters extracted from \figref{fig:ParametersExtraction} (no free parameters). We can see in panel (a) that there is an oscillation of magnitude as a function of $\theta$. This is due to imperfections in our AWG system. When the phase of our AWG system is set to be about 45, 135, 225, and 315 degrees, it outputs a higher power, resulting in the small deviations between theory and experiment.
(e) Linecuts of (c) and (d) with $\theta = \{ 90, 180, 360 \}$ degrees. The difference between these linecuts shows that we can control the loading process by tuning the phase shift $\theta$.
\label{fig:FullDataSlicesPhase}}
\end{figure}

In this section, we present the full data from the phase-shaping experiments discussed in connection with Fig.~4 in the main text. Figure~\ref{fig:FullDataSlicesNum} shows the magnitude of the reflected field using the same $\theta = \pi$ but with $M$ ranging from 1 to 50. Figure~\ref{fig:FullDataSlicesPhase} shows the magnitude of the reflected field using the same $M = 50$ but with $\theta$ varying between 0 and $2 \pi$.

\section{General formalism for single-photon pulse (Fock state)}

The Hamiltonian describing a transmon qubit coupled to a semi-infinitel 1D waveguide can be written as
\be
H = H_S + H_B + H_{\rm int},
\ee
where the bare qubit part is
\be
H_S = \hbar \omega_0 \sp \sm
\ee
and the bare field part is
\be
H_B = \int_0^\infty d \omega \hbar \omega a_\omega^\dag a_\omega,
\ee
with $\sp = \ketbra{e}{g}$ and $\sm = \ketbra{g}{e}$ the ladder operators between the qubit ground state $\ket{g}$ and excited state $\ket{e}$, and $a_\omega^\dag$ ($a_\omega$) the photon creation (annihilation) operator for mode $\omega$. The photonic operators obey the commutation relation $\comm{a_\omega}{a_{\omega'}^\dag} = \delta \mleft( \omega - \omega' \mright)$.

Under the rotating-wave approximation, the interaction Hamiltonian is given by
\be
H_{\rm int} = - i \hbar \int_0^\infty d \omega g \mleft( \omega \mright) \cos \mleft( k_\omega x + \phi_0 \mright) \sm a_\omega^\dag + \text{h.c.},
\label{eq: int. H}
\ee
where $g \mleft( \omega \mright)$ is the coupling strength at frequency $\omega$, the wavenumber is $k_\omega = \omega / v_g$ with $v_g$ the speed of light in the waveguide and h.c.~denotes Hermitian conjugate. Note that the cosine function in $H_{\rm int}$ reflects the formation of a standing wave due to the presence of the mirror at $x = 0$, the end of the transmission line. With the mirror, the phase $\phi_0 = 0$ ($\pi / 2$) corresponds to an anti-node (a node).

The Heisenberg equations of motion of the atomic operators become~\cite{Wang2011, Rag2017}
\bea
\dot{\sigma}_z &=& - \Gamma \mleft( x \mright) \mleft( \sz + I \mright) + 2 \int_0^\infty d \omega g \mleft( \omega \mright) \cos \mleft( k_\omega x \mright) e^{- i \mleft( \omega - \omega_0 \mright) t} \mleft[ \tilde{\sigma}_+ \mleft( t \mright) a_\omega \mleft( 0 \mright) + \text{h.c.} \mright] \label{eq:Heisenberg eqz} \\
\dot{\tilde{\sigma}}_+ &=& - \gamma \mleft( x \mright) \tilde{\sigma}_+ - \int_0^\infty d \omega g \mleft( \omega \mright) \cos \mleft( k_\omega x \mright) e^{i \mleft( \omega - \omega_0 \mright) t}a_\omega^\dag \mleft( 0 \mright) \sz \mleft( t \mright) \label{eq:Heisenberg eqp}
\eea
where we used $\sigma_\pm = \tilde{\sigma}_\pm e^{\pm i \omega t}$ and looked at the slowly varying $\tilde{\sigma}_\pm$. The spontaneous decay rate $T_1^{-1}$ of the qubit is given by $\Gamma (x) = \Gamma \cos^2 \mleft( k_0 x \mright)$ with $\Gamma = 2 \pi g^2 (\omega_0)$. The overall decoherence rate ($T_2^{-1}$) is $\gamma (x) = \Gamma (x) / 2 + \Gamma_\phi$ with the pure dephasing rate $\Gamma_\phi$.

We now further present the formalism for shaping the incident state with precisely one single photon. Such a state can be expressed as
\be
\ket{1_p} = \int_0^\infty d \omega g \mleft( \omega \mright) \cos \mleft( k_\omega x \mright) f \mleft( \omega \mright) a_\omega^\dag \ket{0},
\label{eq: single photon state}
\ee
where $f \mleft( \omega \mright)$ is the spectral distribution function~\cite{Scully1997, Wang2011, Wang2012}, and $\ket{0}$ denotes the multi-mode vacuum state. The initial state of the system is now given by $\ket{\psi_0} = \ket{g, 1_p}$, corresponding to the qubit in its ground state and the incident single-photon state. The evolution of the atomic variables can be determined by the dynamical equations
\bea
\brakket{g, 1_p}{\dot{\sigma}_z}{g, 1_p} &=& - \Gamma \mleft( x \mright) \mleft( \brakket{g, 1_p}{\sz}{g, 1_p} + 1 \mright) + 2 \sqrt{\Gamma \mleft( x \mright)} \mleft( \xi \mleft(t \mright) \brakket{g, 1_p}{\tilde{\sigma}_+}{g, 0} + \xi^* \mleft( t \mright) \brakket{g, 0}{\tilde{\sigma}_-}{g, 1_p} \mright), \label{eq:dynamical eqz} \\
\brakket{g, 1_p}{\dot{\tilde{\sigma}}_+}{g, 0} &=& - \gamma \mleft( x \mright) \brakket{g, 1_p}{\tilde{\sigma}_+}{g, 0} + \sqrt{\Gamma \mleft( x \mright)} \xi^* \mleft( t \mright) \label{eq:dynamical eqp}
\eea
with
\be
\label{eq: input pulse}
\xi \mleft( t \mright) = \frac{\sqrt{\Gamma \mleft( x \mright)}}{2 \pi} \int_0^\infty f \mleft( \omega \mright) e^{- i \mleft( \omega - \omega_0 \mright) t} d \omega
\ee
the temporal waveform of the incident pulse properly normalized by $\int_{-\infty}^\infty \abssq{\xi \mleft( t \mright)} d t = 1$.


\section{Output field and loading efficiency}

In this section, we summarize the derivation of the efficiency for ``loading a photon'' by an incident pulse in the semi-infinite-waveguide architecture. Following the input-output formalism~\cite{Collett1984, Lalumiere2013},
\be
\label{eq: input-output relation}
a_{\rm out} \mleft( t \mright) = a_{\rm in} \mleft( t \mright) - \sqrt{\Gamma \mleft( x \mright)} \tilde{\sigma}_- \mleft( t \mright),
\ee
where the input and output operators are given by
\bea
a_{\rm in} \mleft( t \mright) &=& \frac{1}{\sqrt{2 \pi}} \int_0^\infty d \omega a_\omega \mleft( 0 \mright) e^{- i \mleft( \omega - \omega_0 \mright) t}, \label{eq: input operator} \\
a_{\rm out} \mleft( t \mright) &=& \frac{1}{\sqrt{2 \pi}} \int_0^\infty d \omega a_\omega \mleft( t_f \mright) e^{- i \mleft( \omega - \omega_0 \mright) t} e^{i \omega t_f}, \label{eq: output operator}
\eea
we find the expressions for the input and output signals in terms of the photon flux:
\bea
F_{\rm in} \mleft( t \mright) &=& \expec{a_{\rm in}^\dag \mleft( t \mright) a_{\rm in} \mleft( t \mright)}, \label{eq: input photon flux} \\
F_{\rm out} \mleft( t \mright) &=& \expec{a_{\rm out}^\dag \mleft( t \mright) a_{\rm out} \mleft( t \mright)} = F_{\rm in} \mleft( t \mright) + \Gamma \mleft( x \mright) \expec{\tilde{\sigma}_+ \tilde{\sigma}_-} - \sqrt{\Gamma \mleft( x \mright)} \mleft( \expec{\tilde{\sigma}_+ a_{\rm in}} + \expec{a_{\rm in}^\dag \tilde{\sigma}_-} \mright). \label{eq: output photon flux}
\eea
The loading efficiency is then defined as
\be
\label{eq: loading efficiency}
\eta = \frac{\int_{t_0}^\infty F_{\rm out} \mleft( t \mright) d t}{\int_{-\infty}^\infty F_{\rm in} \mleft( t \mright) d t},
\ee
where $t_0$ refers to the time at which the loading is considered to have ended. Unlike the coherent-state cases, this definition of efficiency captures the contribution of incoherent emission and directly reflects the excitation of the qubit. In our experiments with exponentially rising input pulses, $t_0$ is simply the time when the pulse is turned off: $t_0 = 0$. Note that this efficiency is nothing but the ratio of the emitted energy to the incident energy. In the following, we look at the case with a single-photon Fock state being fed in. The efficiency can be numerically computed by solving the dynamical equations in Eqs.~(\ref{eq:dynamical eqz})-(\ref{eq:dynamical eqp}).


\section{Efficiency of single-photon loading}

Now we focus on the single-photon Fock state $\ket{1_p}$ as the incident state and look at the loading efficiency. Here the state is also shaped into the exponentially rising waveform $\xi \mleft( t \mright) = \sqrt{2 / \tau} e^{t / \tau} \Theta (-t)$ characterized by the time constant $\tau$, where
\be
\Theta(t) = \begin{cases}
0 & t < 0 \\
1 & t \geq 0
\end{cases}
\ee
is the Heaviside step function. The qubit is placed at one of the antinodes $x = 2 n \pi / k_0$ with $k_0 = \omega_0 / v_g$ and $n$ some specific integer. Following the approach discussed in the preceding sections, we obtain the input flux
\be
F_{\rm in} (t) = \frac{2}{\tau} e^{2 t / \tau} \Theta (-t)
\ee
and the output flux
\be
\label{eq:Fout}
F_{\rm out} (t) = \begin{cases}
\mleft[ 1 - \frac{4 \Gamma}{\tau \mleft( \gamma + 1 / \tau \mright) \mleft( \Gamma + 2 / \tau \mright)} \mright] \frac{2 e^{2 t / \tau}}{\tau} & t < 0 \\
\frac{4 \Gamma^2}{\tau \mleft( \gamma + 1 / \tau \mright) \mleft( \Gamma + 2 / \tau \mright)} e^{- \Gamma t} & t > 0
\end{cases}
\ee
with $\gamma = \Gamma / 2 + \Gamma_\phi$. The corresponding loading efficiency is
\be
\eta (\tau) = \frac{4 \Gamma}{\tau \mleft( \gamma + 1 / \tau \mright) \mleft( \Gamma + 2 / \tau \mright)}.
\ee
It is obvious that, for $t < 0$, the output signal is proportional to the input one, with the second term inside the square bracket of \eqref{eq:Fout} accounting for the scattered contribution after interacting with the qubit. During this loading stage, any output flux is unwanted for the purpose of most efficiently loading a photon. The perfect cancellation from destructive interference between the input and scattered signals occurs only when $\tau = 2 / \Gamma$ and $\Gamma_\phi = 0$. Indeed, this gives $\eta = 1$, corresponding to ideal time reversal of spontaneous emission. For general cases with finite pure dephasing, the time constant yielding the highest efficiency is given by $\tau_{\rm opt} = \sqrt{2 / (\gamma \Gamma)}$, which differs slightly from the coherent case discussed in the main text, where $\tau_{\rm opt} = 1 / \gamma$.

\begin{figure}
\begin{centering}
\includegraphics[width=\linewidth]{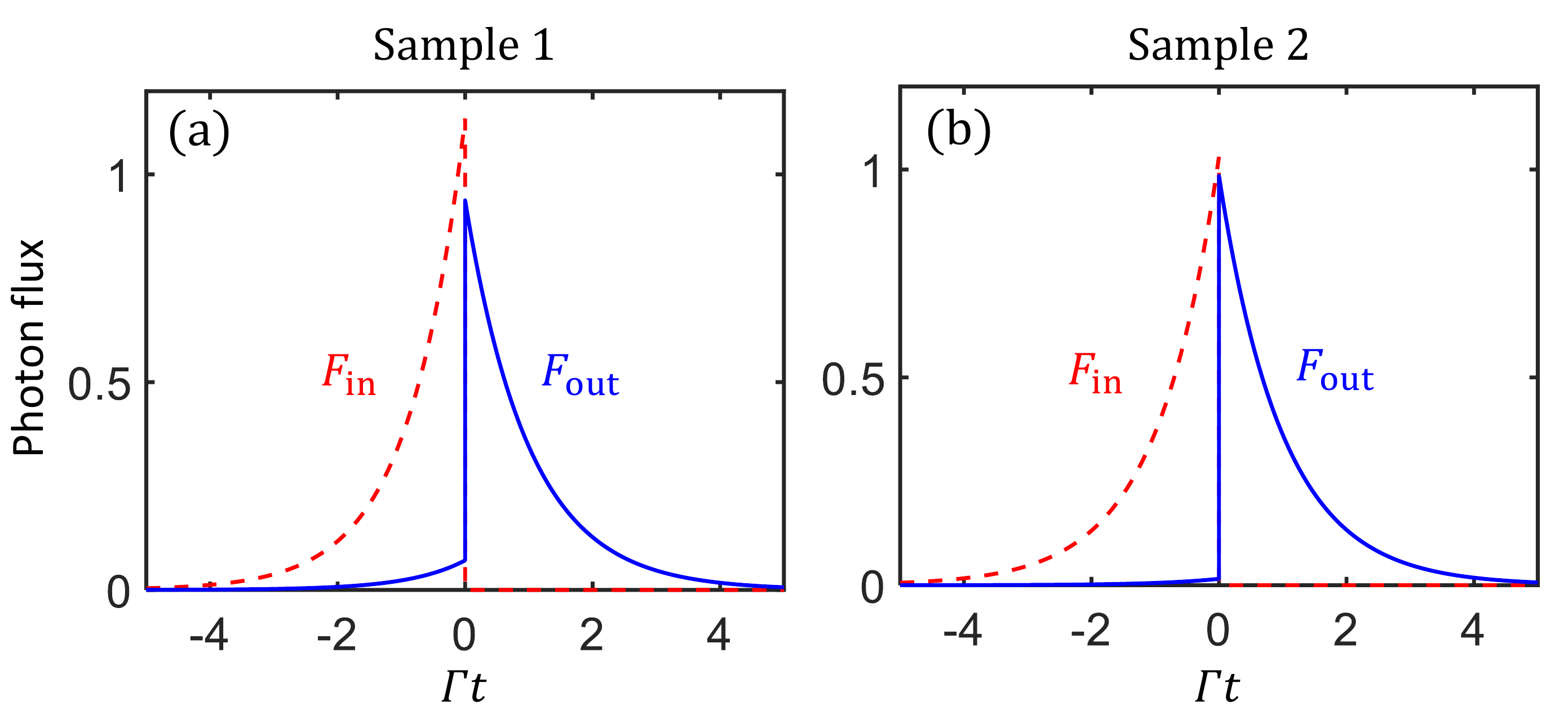}
\par\end{centering}
\caption{Input and output photon flux as functions of time.
(a) and (b) correspond to Samples 1 and 2 with pure dephasing rates $\Gamma_\phi = 0.067 \Gamma$ and $\Gamma_\phi = 0.0152 \Gamma$, respectively. The red dashed curves represent the exponentially rising incident flux $F_{\rm in} (t)$ with $\tau_{\rm opt} = \sqrt{2 / (\gamma \Gamma)}$ and the blue solid curves the resulting output flux $F_{\rm out} (t)$. The fluxes $F_{\rm in/out}$ are given in units of $\Gamma^{-1}$.
\label{fig: input_output_pulse}}
\end{figure}

In the main text, we also discussed two samples in the experiments with different pure dephasing rates. Here we show the results that would be achieved if the incident pulse was a single-photon Fock state with exponentially rising waveform. Figure~\ref{fig: input_output_pulse} shows the instantaneous input and output photon flux for $\tau = \tau_{\rm opt}$. In the first sample, with $\Gamma_\phi / \Gamma \approx \unit[7]{\%}$, we find that there is an observable $F_{\rm out}$ causing excitation loss during the loading stage ($t < 0$) resulting in a loading efficiency $\eta \approx \unit[93.8]{\%}$. This loss is suppressed as $\Gamma_\phi / \Gamma$ decreases, as in the second sample, where $\Gamma_\phi / \Gamma \approx \unit[1.5]{\%}$. The almost perfect symmetry with respect to $t = 0$ in $F_{\rm in/out}$ implies near unit efficiency: $\eta \approx \unit[98.5]{\%}$. This cannot be matched in the coherent cases with mean photon number equal to 1, since such states also contain a non-negligible two-photon component.

\begin{figure}
\begin{centering}
\includegraphics[width=\linewidth]{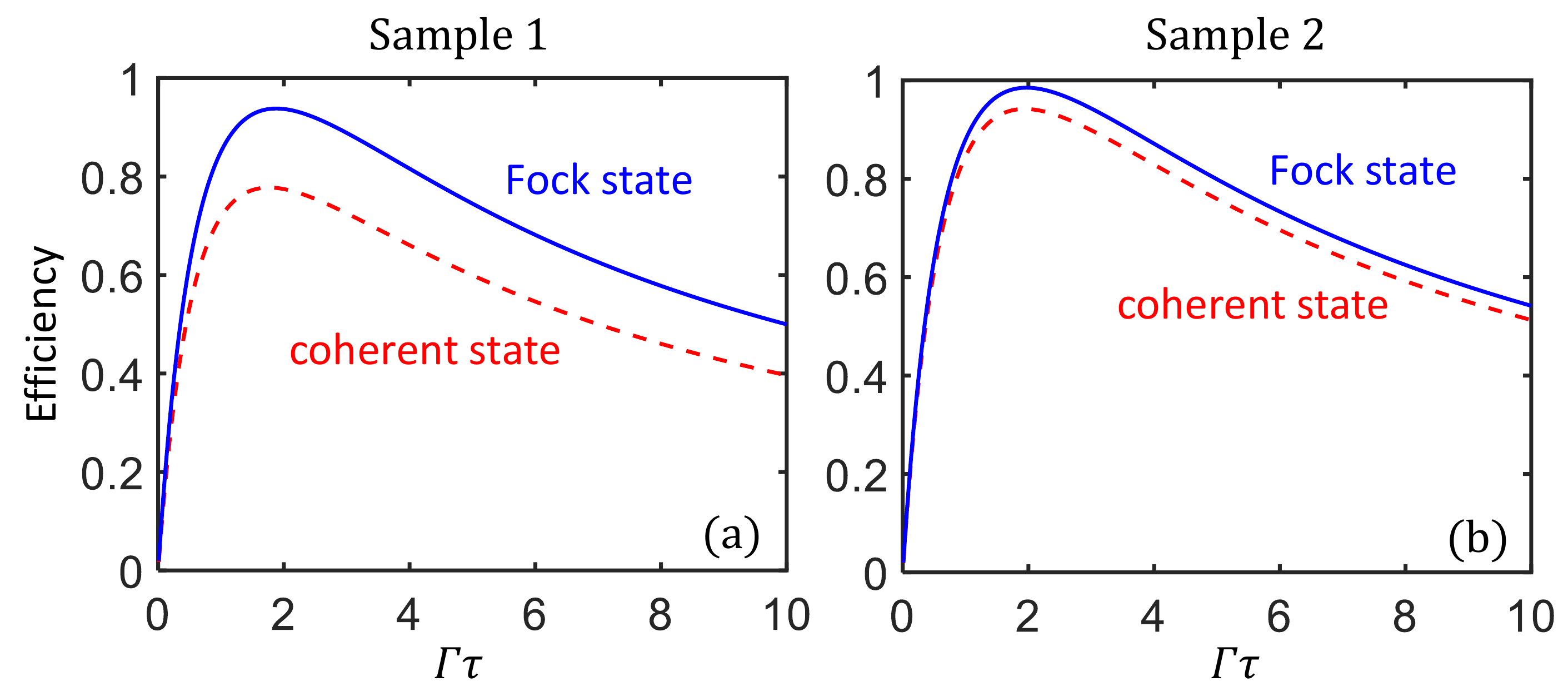}
\par\end{centering}
\caption{Loading efficiency as a function of the time constant $\tau$ of the exponentially rising input pulse.
(a) and (b) correspond to Samples 1 and 2. The efficiency for the Fock-state input is plotted as blue solid curves and the efficiency for the weak coherent-state input is shown as red dashed curves.
\label{fig: Loading efficiency tau}}
\end{figure}

\begin{figure}
\begin{centering}
\includegraphics[width=\linewidth]{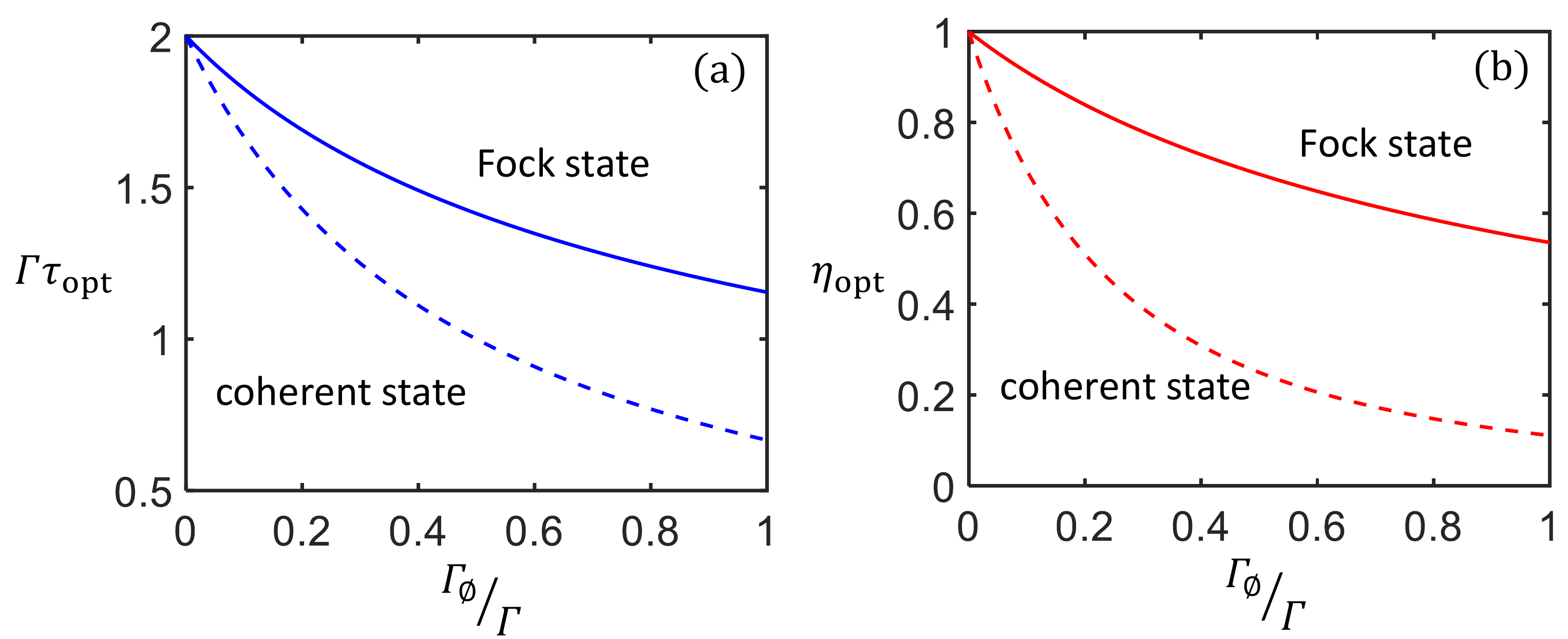}
\par\end{centering}
\caption{Effect of pure dephasing.
(a) Optimal time constant as a function of the pure dephasing rate for Fock-state input (solid curve) and weak coherent input (dashed curve).
(b) The corresponding loading efficiency as a function of the pure dephasing rate for Fock-state input (solid curve) and weak coherent input (dashed curve).
\label{fig: maximum loading efficiency}}
\end{figure}

Figure~\ref{fig: Loading efficiency tau} displays the loading efficiency as a function of the characteristic time constant for the two samples. It can be observed that the loading efficiency is more robust against pure dephasing for Fock state input than for coherent input in general. Interestingly, although defined slightly differently, the loading efficiency shows almost identical dependence on the characteristic time constant in the coherent and single-photon-Fock-state cases as $\Gamma_\phi \rightarrow 0$, reaching $\eta_{\rm opt} \rightarrow 1$ at $\tau_{\rm opt} = 2 / \Gamma$. Figure~\ref{fig: maximum loading efficiency} shows the effect of finite pure dephasing rates, which unfortunately always degrade the loading efficiency. Note that the optimal time constant also is shortened as $\Gamma_\phi$ increases, which can be somewhat expected qualitatively: since the linewidth of the qubit is broadened, the needed incident pulse duration must be shorter in order to have better overlap in the spectral distribution.

\bibliography{LoadingPhotonRefs}